\documentclass[symmetry,article,submit,moreauthors,pdftex]{mdpi} 

\firstpage{1} 
\pubvolume{1}
\issuenum{1}
\articlenumber{0}
\pubyear{2021}
\copyrightyear{2020}
\datereceived{} 
\dateaccepted{} 
\datepublished{} 
\hreflink{https://doi.org/}

\usepackage{overpic}
\usepackage{mathtools}

\Title{Proton Electromagnetic Form Factors in the Time-like Region through the Scan Technique}

\TitleCitation{Proton Electromagnetic Form Factors in the Time-like Region through the Scan Technique}

\Author{L.~Xia$\orcidA{}^{1,2*}$, Ch.~Rosner$\orcidB{}^{3*}$, Y.~D.~Wang$\orcidC{}^{4*}$, X.~R.~Zhou$\orcidD{}^{1,2*}$, F.~E.~Maas$\orcidE{}^{3*}$, R.~Baldini Ferroli$\orcidF{}^{5*}$, H.~M.~Hu$\orcidG{}^{6,7*}$, G.~S.~Huang$\orcidH{}^{1,2*}$}

\AuthorNames{L.~Xia, Ch.~Rosner, Y.~D.~Wang, X.~R.~Zhou, F.~E.~Maas, R.~Baldini~Ferroli, H.~M.~Hu and G.~S.~Huang}

\AuthorCitation{Xia, L.; Rosner, Ch.; Wang, Y.~D.; Maas, F.~E.; Ferroli, R.~Baldini; Hu, H.~M; Huang, G.~S.}

\address{
$^{1}$ University of Science and Technology of China, Hefei 230026, People's Republic of China\\
$^{2}$ State Key Laboratory of Particle Detection and Electronics, Hefei 230026, People's Republic of China\\
$^{3}$ Helmholtz Institute Mainz, Staudinger Weg 18, D-55099 Mainz, Germany\\
$^{4}$ North China Electric Power University, Beijing 102206, People's Republic of China\\
$^{5}$ INFN Laboratori Nazionali di Frascati, I-00044, Frascati, Italy\\
$^{6}$ Institute of High Energy Physics, Beijing 100049, People's Republic of China\\
$^{7}$ University of Chinese Academy of Sciences, Beijing 100049, People's Republic of China\\}
\corres{Correspondence: xial@ustc.edu.cn (Xia, L), chrosner@uni-mainz.de (Rosner, Ch.), wangyadi@ncepu.edu.cn (Y.~D.~Wang), zxrong@ustc.edu.cn (X.~R.~Zhou), maas@him.uni-mainz.de (F.~E.~Maas), Rinaldo.Baldini@lnf.infn.it (R.~Baldini~Ferroli), huhm@ihep.ac.cn (H.~M.~Hu), hgs@ustc.edu.cn (G.~S.~Huang)}

\abstract{For over 100 years, scientists have investigated the properties of the proton, which is one of the most abundant components of visible matter in the universe. 
Nevertheless, researchers do not fully understand many details about its internal structure and dynamics. 
Time-like electromagnetic form factors are one of the observable quantities that can help us achieve a deeper understanding.
In this review article, we present an overview of the current experimental status in this field, consisting of measurements of the time-like reactions $e^{+}e^{-}\to p\bar{p}$, $p\bar{p}\to e^{+}e^{-}$, and future measurements of $p\bar{p}\to \mu^{+}\mu^{-}$.
A focus is put on recent high precision results of the reaction $e^{+}e^{-}\to p\bar{p}$ that have been obtained after analyzing 688.5~pb$^{-1}$ of data taken at the BESIII experiment. 
They are compared and put into perspective to results from previous measurements in this channel. 
We discuss the channels $p\bar{p}\to e^{+}e^{-}$ and $p\bar{p}\to\mu^{+}\mu^{-}$ in terms of the few existing as well as future measurements, which the PANDA experiment will perform.
Finally, we review several new theoretical models and phenomenological approaches inspired by the BESIII high precision results and then discuss their implications for a deeper understanding of the proton’s structure and inner dynamics.
}


\begin{document}
\section{Introduction}

The proton, along with its partner nucleon, the neutron, makes up more than 99.9\% of the visible matter in the universe. The internal structure and dynamics of the proton, which originate from the strong interaction, has therefore been the target of intense investigations over the past 100 years since the discovery of the proton. Despite that, many questions are still not answered satisfactorily, recent examples being the proton radius puzzle, the proton spin crisis, or the origin of the proton mass. A detailed theoretical description of the internal proton structure and its constituents' dynamics is made difficult by the non-perturbative nature of the underlying theory, quantum chromodynamics (QCD), within the energy regime of the nucleon. Therefore, precise knowledge of one of the most simple observables that parametrize the proton's structure and dynamics, the electromagnetic (EM) form factors (FFs), is crucial for understanding the proton structure. In turn, these quantities also provide a perfect testing ground for our understanding of QCD.

\begin{figure}[htbp]
\begin{center}
\begin{overpic}[width=0.36\textwidth]{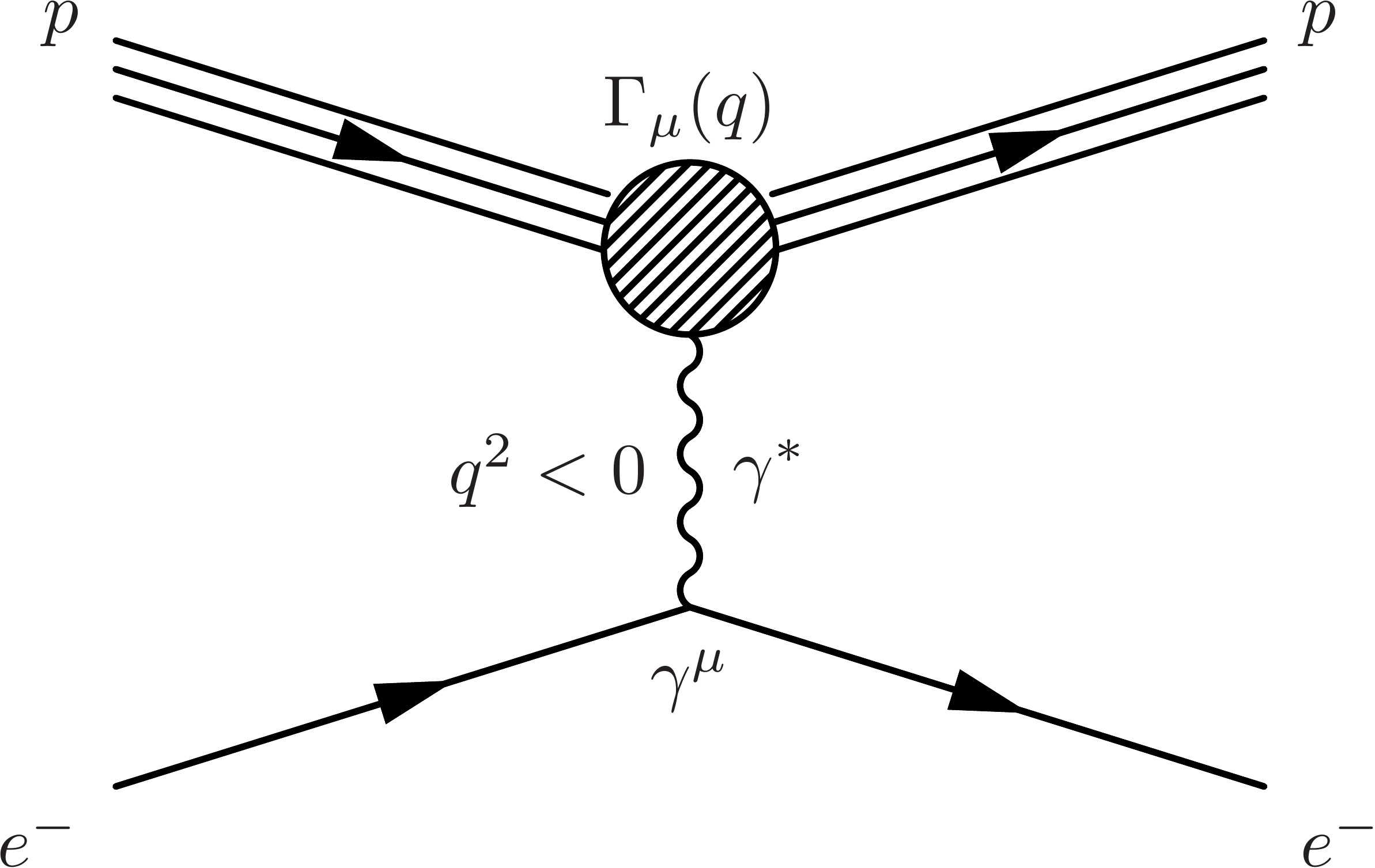}
\put(45,0){(a)}
\end{overpic}
\begin{overpic}[width=0.36\textwidth]{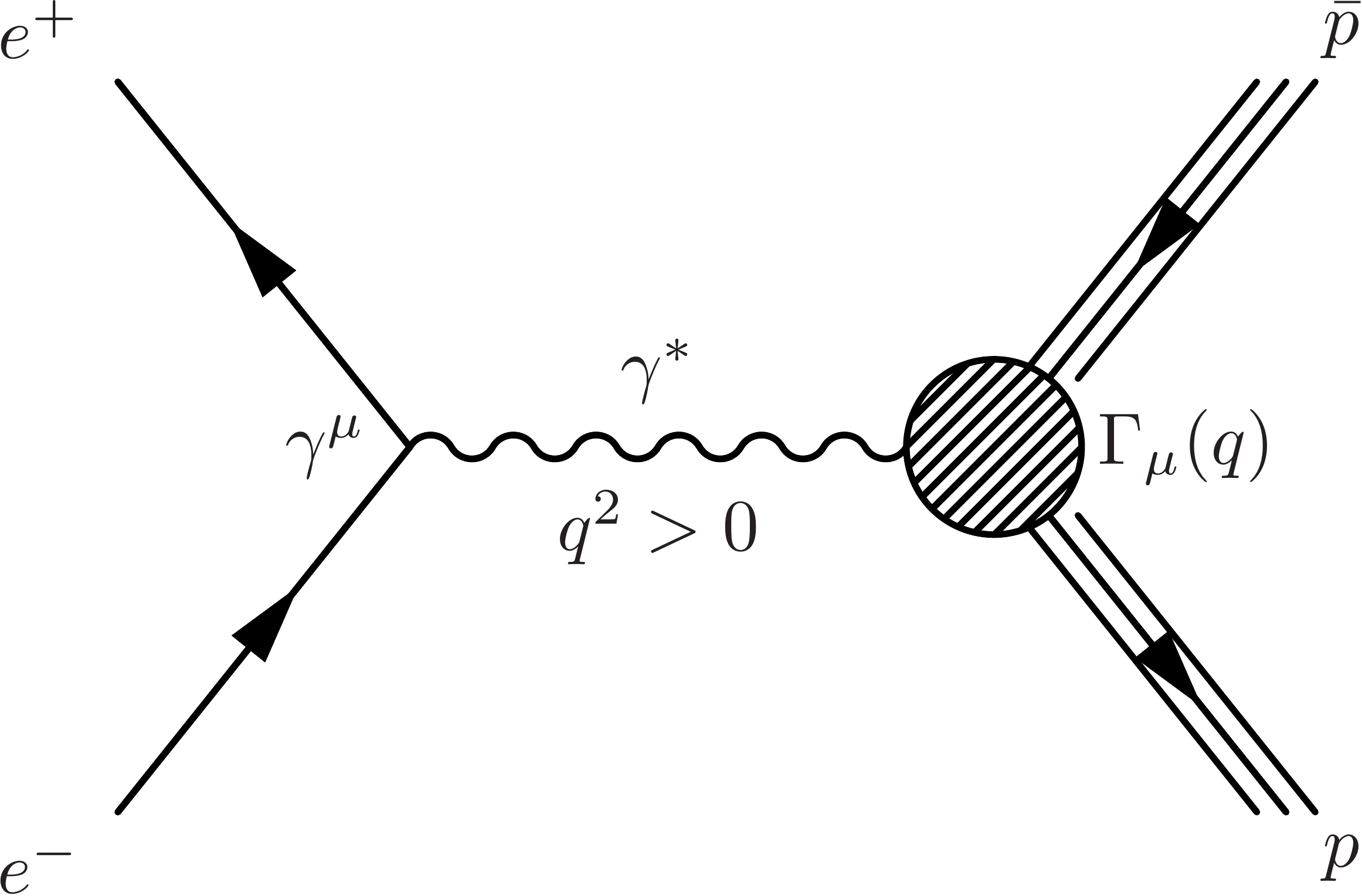}
\put(45,0){(b)}
\end{overpic}
\caption{Lowest-order Feynman diagrams for elastic electron-proton scattering $e^{-}p\rightarrow e^{-}p$ (a), and for the annihilation process $e^{+}e^{-}\rightarrow p\bar{p}$ (b).}
\label{fig_Feynman_SL_TL}
\end{center}
\end{figure}

The discovery of the proton dates back to the earliest investigations of the atomic structure. 
W.~Wien and J.~J.~Thomson identified a positively charged particle with a mass equal to the hydrogen atom when studying streams of ionized gaseous atoms and molecules whose electrons had been stripped.
E.~Rutherford indicated that the nitrogen under $\alpha$ particle bombardment ejects what appear to be hydrogen nuclei in 1919.
Subsequently, in 1920, E.~Rutherford had accepted the hydrogen nucleus as an elementary particle, denominating its proton.
In 1937, O~Stern observed an anomalous magnetic moment of nucleons ($\mu_{p}=2.79\mu_{N}$, $\mu_{n}=-1.91\mu_{N}$), where $\mu_{N}=\frac{e\hbar}{2m_{N}c}$ is the nuclear magneton,
while theoretically, the magnetic moment of point-like proton and neutron is $\mu_{N}$ and 0, respectively.
Furthermore, it is the most direct evidence that nucleons are not point-like particles~\cite{Stern}.
Afterward, R.~Hofstadter established an apparatus for discovering nuclei's internal structure by elastic scattering of electrons and protons (see Figure~\ref{fig_Feynman_SL_TL}~(a)).
In the 1950s, R.~Hofstadter investigated the charge distribution in atomic nuclei and the charge and magnetic moment distributions in the proton and neutron.
Nuclei were thereby proven not to be homogeneous but to have internal structures~\cite{Hofstadter}.
After that, J.~I.~Friedman, H.~W.~Kendall, and R.~E.~Taylor investigated deep inelastic scattering of electrons on protons and bound neutrons (also see Figure~\ref{fig_Feynman_SL_TL}~(a)).
This research has been of essential significance for the development of the quark model in particle physics~\cite{Friedman}.

Until today, the research of the internal structures of protons is one of the hottest topics in nuclear physics.
The proton-spin crisis and proton-radius puzzle are the most critical issues of the research.
In 2017, $\chi$QCD Collaboration reported that the gluon helicity contributes to half of the total proton spin by Lattice QCD~\cite{YiboYang}.
Afterwards, Jefferson Laboratory (PRad) reported a proton radius of $r_{p}=0.831\pm 0.007({\text{stat}})\pm 0.012({\text{syst}})$~\cite{radius}, which supports the value found by two previous muonic hydrogen experiments~\cite{radius_old,radiusMuon2}.

Nucleon EM FFs parametrize the difference between a point-like photon-nucleon vertex to one that considers the internal structure of the nucleon. 
For spin 1/2 particles, such as proton and neutron, this structure is encoded in two FFs, the Dirac FF ($F_{1}$) and the Pauli FF ($F_{2}$), where the first describes the difference to a point-like charge distribution and the second to a point-like magnetization distribution. More commonly used than these Dirac- and Pauli FFs are the so-called Sachs FFs $G_{E}$ and $G_{M}$, which are simple linear combinations of the former. Proton EM FFs can be explored in two kinematic regions, the spacelike (SL, momentum transfer $q^{2}<0$) and timelike (TL, momentum transfer $q^{2}>0$) region. The former region has been extensively investigated through electron-proton scattering experiments (see Figure~\ref{fig_Feynman_SL_TL}~(a)) since the 1950s, and consequently, proton TL EM FFs are known with low uncertainties of the order of a few percent. Experimental data on the TL region, commonly obtained by investigating annihilation reactions (Figure~\ref{fig_Feynman_SL_TL}~(b)), has been much more scarce for a long time. Only within the last two decades, more experiments have provided data within this region, and only very recent high luminosity measurements have brought the precision up to par with the SL region.
FFs enter explicitly in the coupling of a virtual photon with the hadron electromagnetic current. Their measurements can be directly compared to hadron models and thereby provide constraints to the description of the internal structure of hadrons~\cite{radius_old}.
The extension of the proton EM FFs to the TL region opens further possibilities to investigate peculiarities in the structure of the proton. While the information contained in the TL FFs has a less intuitive interpretation compared to the electric and magnetic distribution densities of the nucleon deduced from SL FFs, they play a crucial role in understanding the long-range behavior of strong interactions. Moreover, proton EM FFs can be used to make theoretical predictions for the behavior of other baryons' FFs, e.g. the neutrons~\cite{NeutronPrediction}.

In terms of $G_{E}$ and $G_{M}$, the cross section of the process $e^{+}e^{-}\xleftrightarrow{} p\bar{p}$ reads:
\begin{eqnarray}
\label{eq_FF1}
\sigma_{e^{+}e^{-}\xleftrightarrow{} p\bar{p}}(s)=\frac{4\pi\alpha^{2}\beta C}{3s}\left[|G_{M}(s)^{2}|+\frac{2m_{p}^{2}}{s}|G_{E}(s)^{2}|\right],
\end{eqnarray}
where $\alpha$ is the fine structure constant, $\beta=\sqrt{1-4m_{p}^{2}/s}$ is the proton velocity, $m_{p}$ is the proton mass, and \textit{C} is the coulomb enhancement factor~\cite{Rinaldo_coul1}.
From this integrated cross section, a so called effective FF ($G_{\text{eff}}$) can be deduced under the assumption of $|G_{E}(s)|=|G_{M}(s)|$:
\begin{eqnarray}
\label{eq_FF2}
|G_{\text{eff}}(s)|=  \sqrt{\frac{\sigma_{e^{+}e^{-}\xleftrightarrow{} p\bar{p}}(s)}{\frac{4\pi\alpha^2\beta C}{3s}(1+\frac{2m_{p}^{2}}{s})}} =\sqrt{\frac{|G_{M}(s)^{2}|+\frac{2m_{p}^{2}}{s}|G_{E}(s)^{2}|}{1+\frac{2m_{p}^{2}}{s}}}.
\end{eqnarray}
This $|G_{\text{eff}}|$ was mainly used by older experiments with limited statistics and was calculated from the measured cross section using the middle part of the above equation.
More recent experiments have been able to perform a measurement of the differential cross section through an angular analysis in one-photon exchange approximation in the $e^{+}e^{-}$ center-of-mass (c.m.) system, which allows to determine the individual FFs $|G_{E}(s)|$ and $|G_{M}(s)|$:
\begin{eqnarray}
\label{eq_FF3}
    \frac{d\sigma_{p\bar{p}}(s)}{d\Omega}=\frac{\alpha^{2}\beta C}{4s}\left[|G_{M}(s)|^{2}(1+\cos^{2}\theta)+\frac{4m_{p}^{2}}{s}|G_{E}(s)|^{2}\sin^{2}\theta\right],
\end{eqnarray}
where $\theta$ is the polar angle of the outgoing particles.

Various approaches to providing a theoretical description of the proton EM FFs exist, among them approaches based on modified versions of the old, yet still relevant, vector meson dominance (VMD) model~\cite{VMDModel}, models based on the unitarity and analyticity of the proton FFs using dispersion theoretical approaches~\cite{DispersionTheory}, descriptions based on a relativistic constituent quark model~\cite{CQM}, and more.
A comprehensive review of all these approaches is outside the scope of this article. Instead, a focus is put on a review of the experimental results for the proton TL FFs, especially from the most recent high luminosity BESIII energy scan, and the current, primarily phenomenological theoretical approaches to describe the newly observed features in the effective TL proton FF as well as the individual electric and magnetic FF.

\section{Scan method}

\begin{figure}[htbp]
\begin{center}
\begin{overpic}[width=0.23\textwidth]{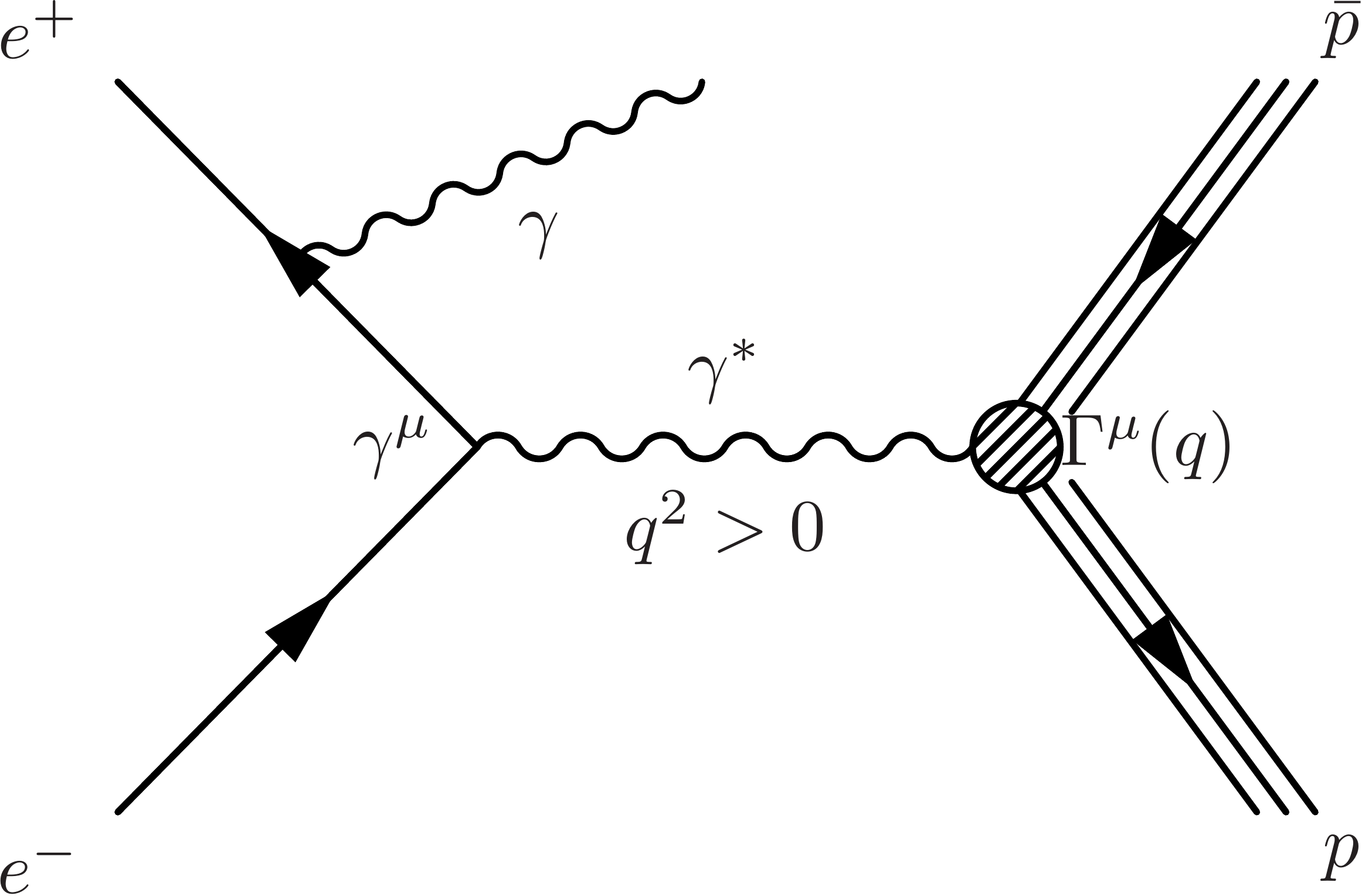}
\put(45,0){(a)}
\end{overpic}
\begin{overpic}[width=0.23\textwidth]{./feyngraph_TL_scan.pdf}
\put(45,0){(b)}
\end{overpic}
\begin{overpic}[width=0.23\textwidth]{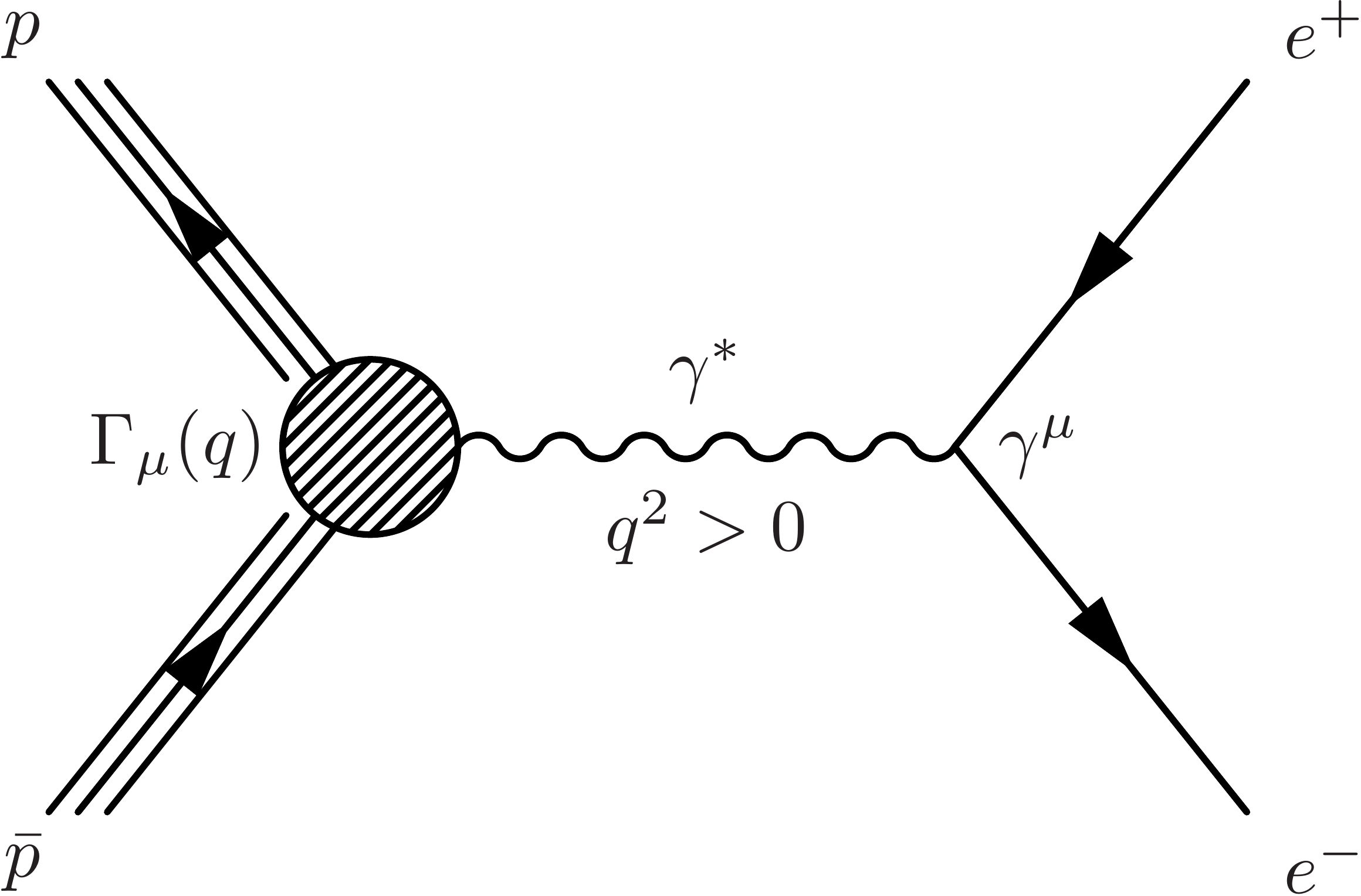}
\put(45,0){(c)}
\end{overpic}
\caption{Feynman diagrams for the ISR (a) and born (b) process of $e^{+}e^{-}\rightarrow p\bar{p}$, as well as the time-reverse process $ p\bar{p}\rightarrow e^{+}e^{-}$ (c).}
\label{fig_Feynman}
\end{center}
\end{figure}

Measuring nucleon EM FFs over a wide kinematical range is possible through two different methods, the so-called initial state radiation (ISR) and scan methods. The ISR method uses collider data at a fixed c.m. energy and analyses events where a photon is emitted from the initial state (see Figure~\ref{fig_Feynman} (a)), thus reducing the momentum transfer $q^{2}$ of the process. This allows measuring FFs at $q^{2}$ values from threshold up to the fixed c.m.~energy of the collider. For the scan method, this is achieved by taking energy scan data sets, where the c.m.~energy of the collider is systematically varied. Within this paper, only former results, as well as future experiments employing the scan method, are reviewed, whereas an overview of results with the ISR method can be found in Ref~\cite{AlaaDexu}. Further reviews about nucleon EM FFs in the TL region in general and other baryon FFs can be found in Ref.~\cite{DenigReview} and Ref.~\cite{GuangshunReview}, respectively.

\section{Overview of Timelike Electromagnetic Form Factor measurements for the Proton}

\subsection{Previous measurements of the proton EM FFs}

Previous measurements of the proton EM FFs using the scan method have been performed with the processes $e^{+}e^{-}\rightarrow p\bar{p}$ as well as the time-reverse process, $p\bar{p}\rightarrow e^{+}e^{-}$ (see Figure~\ref{fig_Feynman}~(b) and (c)).
While processes where muons replace the electron and positron, are possible, no FF measurements have been performed employing these channels yet.
So far, such measurements were infeasible either due to the lack of $\mu^{+}\mu^{-}$ colliders, or in the time-reversed case due to a combination of low statistics and high background contamination, and the same is true for tauons.
Proposals for measurements in the muon channel will be discussed in the next section.

The earliest attempts to measure proton EM FFs in the time-like region have been performed in the 1970s, in case of electron-positron annihilation at the ADONE collider in Frascati~\cite{Adone} and in case of $p\bar{p}$ annihilation reactions with the Proton Synchrotron (PS) at CERN~\cite{MSTColl2}.
In the 1980s and 90s, measurements in electron-positron annihilation were continued at these facilities with the FENICE experiment~\cite{introFENICE} at ADONE as well as the DM1~\cite{introDM1} and DM2~\cite{introDM2} experiments at the Orsay colliding beam facility (DCI) in Orsay. Fenice measured at 5 energy scan points between 1.90 and 2.44~GeV with a total integrated luminosity ($\mathcal{L}_{\text{int}}$) of 0.36~pb$^{-1}$, whereas DM1 and its successor DM2 both did scans with an integrated luminosity of about 0.4~pb$^{-1}$ between 1.925 and 2.226~GeV in 4 and 6 energy steps, respectively. The time-reversed channel was further explored with the PS170~\cite{introPS170} experiment, using the $\bar{p}$ beam available at the Low Energy Antiproton Ring (LEAR) at CERN, in a series of scan measurements performed between 1991 and 1994 with momentum transfer from the proton-antiproton production threshold up to 2.05~GeV.
Additional measurements using this reaction at higher momentum transfer were also performed at the E760~\cite{introE760} and E835~\cite{introE835} experiments at Fermilab, in case of E760 in 3 steps between 2.98 and 3.6~GeV ($\mathcal{L}_{\text{int}}=26.5$~pb$^{-1}$) and in case of E835 in a total of 10 steps between 2.97 and 4.29~GeV ($\mathcal{L}_{\text{int}}=194.5$~pb$^{-1}$).
In the 2000s, the first high precision, high luminosity measurements started in the electron-positron channel in the form of an extended energy scan between 2.00 and 3.07~GeV performed with the Beijing Spectrometer (BES) at the Beijing electron positron collider (BEPC)~\cite{introBES}, as well as a single high luminosity scan point at 3.671~GeV obtained with the CLEO-c detector at the Cornell Electron Storage Ring (CESR)~\cite{introCLEO}.
The CMD-3 experiment at the VEPP-2000 $e^{+}e^{-}$ collider performed a fine energy scan of the proton TL EM FFs between threshold and 2~GeV, divided into 10 energy points in 2016, and an additional 11 energy points in 2019~\cite{cmd3}.
Finally, the most recent and by far most precise results for the proton time-like EM FFs obtained with the energy scan technique stem from a new high luminosity scan measurement performed in 2015~\cite{xiaorong} and 2020~\cite{BES3scanLXRC} with the BESIII detector at BEPCII. This new data comprises 22 energy scan points between 2.00 and 3.08~GeV, with a total integrated luminosity of 688.5~pb$^{-1}$. The high luminosity of this new data allows a precise determination of the individual FFs $|G_{E}|$ and $|G_{M}|$ as well as their ratio R, whereas most previous experiments have been limited to the determination of $|G_{\text{eff}}|$.

\subsection{Future experimental prospects}

Further measurements of the proton TL EM FFs are necessary to confirm the results obtained by the newest BESIII measurement for the individual FF $|G_{E}|$ and $|G_{M}|$ as well as their ratio.
In addition, more data is necessary to investigate newly found structures that appear both in the effective and the individual FFs. 
An extension of the measured kinematical range for the separate FFs, both towards higher c.m.~energies and the threshold region, is also desirable.

For proton TL EM FF measurements with the scan technique, the most promising upcoming experiment is the PANDA experiment currently under construction at the Facility for Antiproton and Ion Research (FAIR) in Darmstadt.
Located at the High Energy Storage Ring (HESR), which will provide an $\bar{p}$ beam with luminosities up to $10^{32}$~cm$^{-2}$s$^{-1}$ and momenta between 1.5 and 15~GeV/$c$, the PANDA detector will be divided into a target spectrometer and a forward spectrometer. A detailed description of the detector can be found in Ref~\cite{PandaDetector}. 
 
Proposals for proton TL EM FF measurements exist both for the channel $p\bar{p}\rightarrow e^{+}e^{-}$ as well as  $p\bar{p}\rightarrow\mu^{+}\mu^{-}$. 
Measurements, especially of the individual proton TL EM FFs employing the first channel, will be of high interest since here the only existing. Rather an old measurement by PS170 is in disagreement with newer $e^{+}e^{-}\rightarrow p\bar{p}$ results both from the scan and ISR method.
The second channel will be the first access to proton TL EM FFs using leptons other than electrons, which will also allow for tests of lepton universality. 
Feasibility studies exist for both channels: for the electron channel~\cite{PandaFeasibility1}, expected relative uncertainties range between 1\% to 5\% for the FF ratio ($|G_{E}/G_{M}|$) and 1\% to 4\% for $|G_{E}|$ in the already explored $q^{2}$ range up to 3~GeV/$c$ . 
The expected uncertainty for both R and $|G_{E}|$ grows to more than 50\% at 3.73~GeV/$c$. However, this would still be an essential extension of the kinematical range of the measurement of the individual FFs.  
Uncertainties for $|G_{M}|$ would be well below 1\% for all momentum transfers.
In case of the muon channel~\cite{PandaFeasibility2}, uncertainties would range between 3\% to 27\% for $|G_{E}|$, 2\% to 10\% for $|G_{M}|$, and 5\% to 37\% for the $|G_{E}/G_{M}|$ in the energy range between 2.25 and 2.86~GeV. 

\section{Discussion of proton TL EM FF results}
Within this section, the different TL EM FF measurements of the experiments introduced in the last section are compared and put into perspective. The comparison starts with the cross section of the process $e^{+}e^{-}\xleftrightarrow{}p\bar{p}$ in Section~\ref{sec_XsComp}, which was determined by most experiments performing scan measurements of proton FFs. While it is not directly a measurement of proton FFs, the cross section is closely related to the effective FF $|G_{\text{eff}}|$ (see Eq.~(\ref{eq_FF2}), which is discussed in Section~\ref{sec_GeffComp}. Therefore, the discussion in the latter section applies to both the cross section and $|G_{\text{eff}}|$. Finally, the measurements of the FF ratio as well as the individual electric ($|G_{E}|$) and magnetic ($|G_{E}|$) proton FF are compared in Section~\ref{sec_RComp} and~\ref{sec_FFComp}, respectively.
\subsection{Measurements of the cross section of $e^{+}e^{-}\xleftrightarrow{}p\bar{p}$}
\label{sec_XsComp}

\begin{figure}[htbp]
\begin{center}
\centering
\mbox{
  \begin{overpic}[width=6.6cm,height=5.0cm,angle=0]{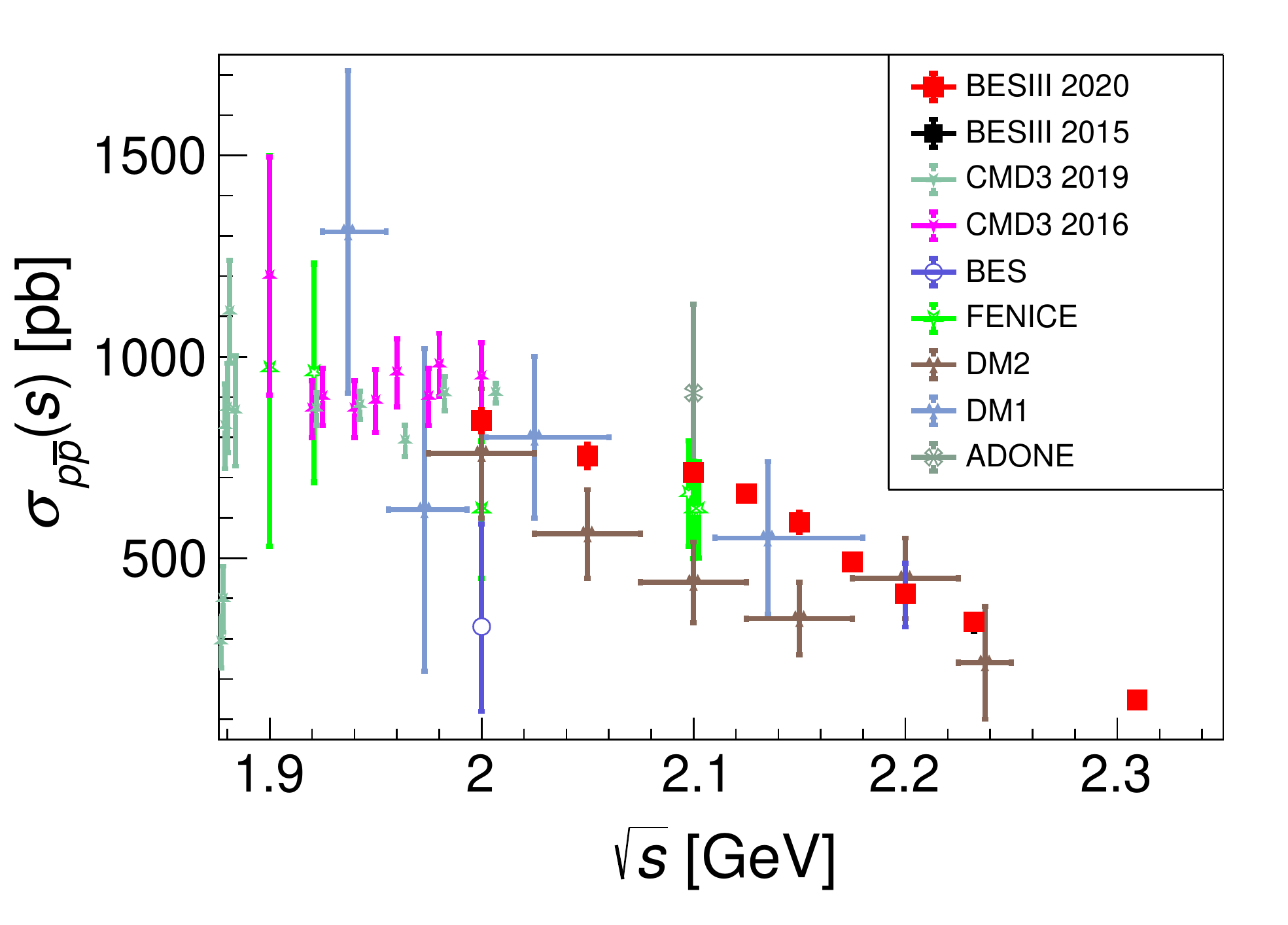}
  \put(0,69){$(a)$}
  \end{overpic}
 \begin{overpic}[width=6.6cm,height=5.0cm,angle=0]{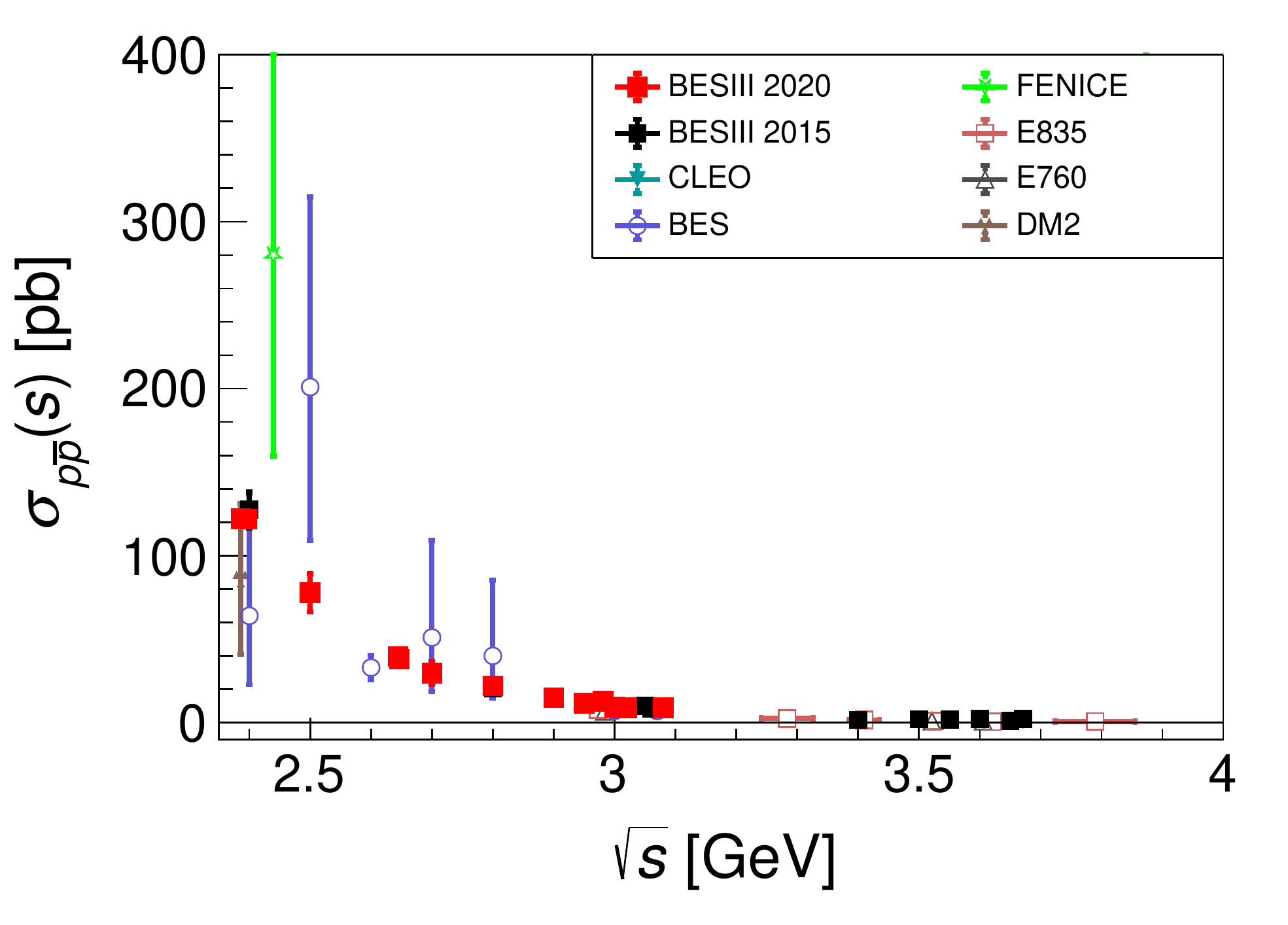}
  \put(0,69){$(b)$}
  \end{overpic}
 }
 \vskip-5.0pt
 \caption{
Comparison of the results for the $e^{+}e^{-} \xleftrightarrow{} p\bar{p}$ cross section using the energy scan strategy in (a) $2m_{p}<\sqrt{s}<2.35$~GeV,  (b) $2.35<\sqrt{s}<4.00$~GeV.
Shown are the published measurements from BESIII~\cite{xiaorong,BES3scanLXRC}, CMD3~\cite{cmd3}, CLEO~\cite{introCLEO}, BES~\cite{introBES}, FENICE~\cite{introFENICE}, E835~\cite{introE835}, E760~\cite{introE760}, DM2~\cite{introDM2}, DM1~\cite{introDM1}, and ADONE~\cite{Adone}.
}
\label{fig_Xsec_result1}
\end{center}
\end{figure}

Measurements of the cross section of $e^{+}e^{-}\xleftrightarrow{}p\bar{p}$ from $e^{+}e^{-}$ colliders employing the scan technique are summarized in Figure~\ref{fig_Xsec_result1}~(a) for c.m.~energies from threshold up to 2.35~GeV, and in  Figure~\ref{fig_Xsec_result1}~(b) for c.m.~energies between 2.35 and 4.00~GeV.

\subsection{Measurements of the effective FF $|G_{\text{eff}}|$ of the proton}
\label{sec_GeffComp}

Measurements of the proton EM FFs of most previous experiments in the TL region were restricted to measuring $|G_{\text{eff}}|$, as shown in Eq.~(\ref{eq_FF2}), due to limited statistics. 
Instead of the assumption of $|G_{E}|=|G_{M}|$, some experiments at high values of $q^{2}$ (e. g. E760 and E835) also measured the magnetic FF $|G_{M}|$ under the assumption that the $|G_{E}|$ contribution is negligible at high energies due to the suppression by a factor of $1/\tau$ (see Eq.~(\ref{eq_FF3})), with $\tau=\frac{s}{4m_p^2}$.

\begin{figure}[htbp]
\begin{center}
\centering
\vskip-0pt
\mbox{
  \begin{overpic}[width=6.6cm,height=5.0cm,angle=0]{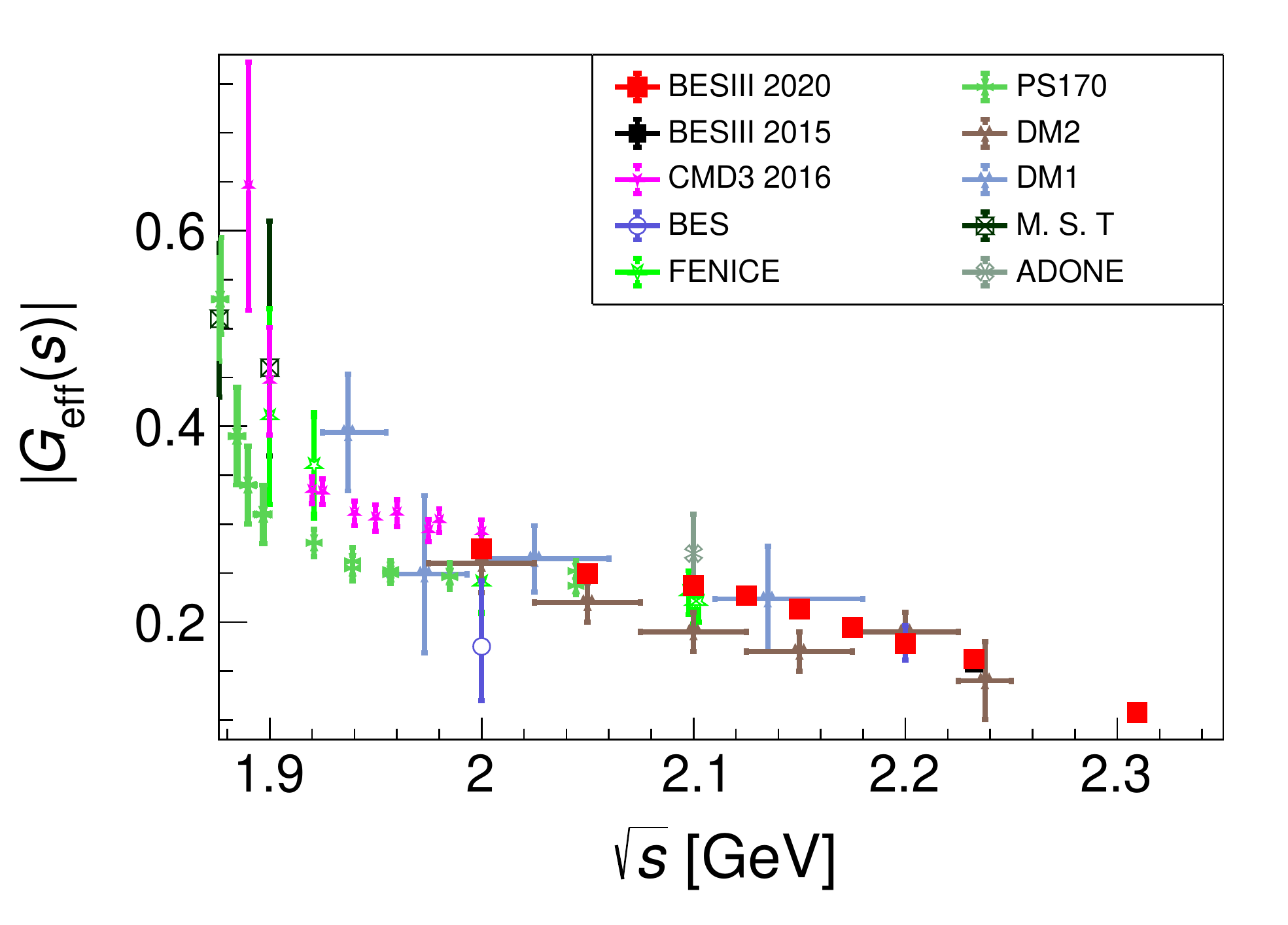}
  \put(0,69){$(a)$}
  \end{overpic}
 \begin{overpic}[width=6.6cm,height=5.0cm,angle=0]{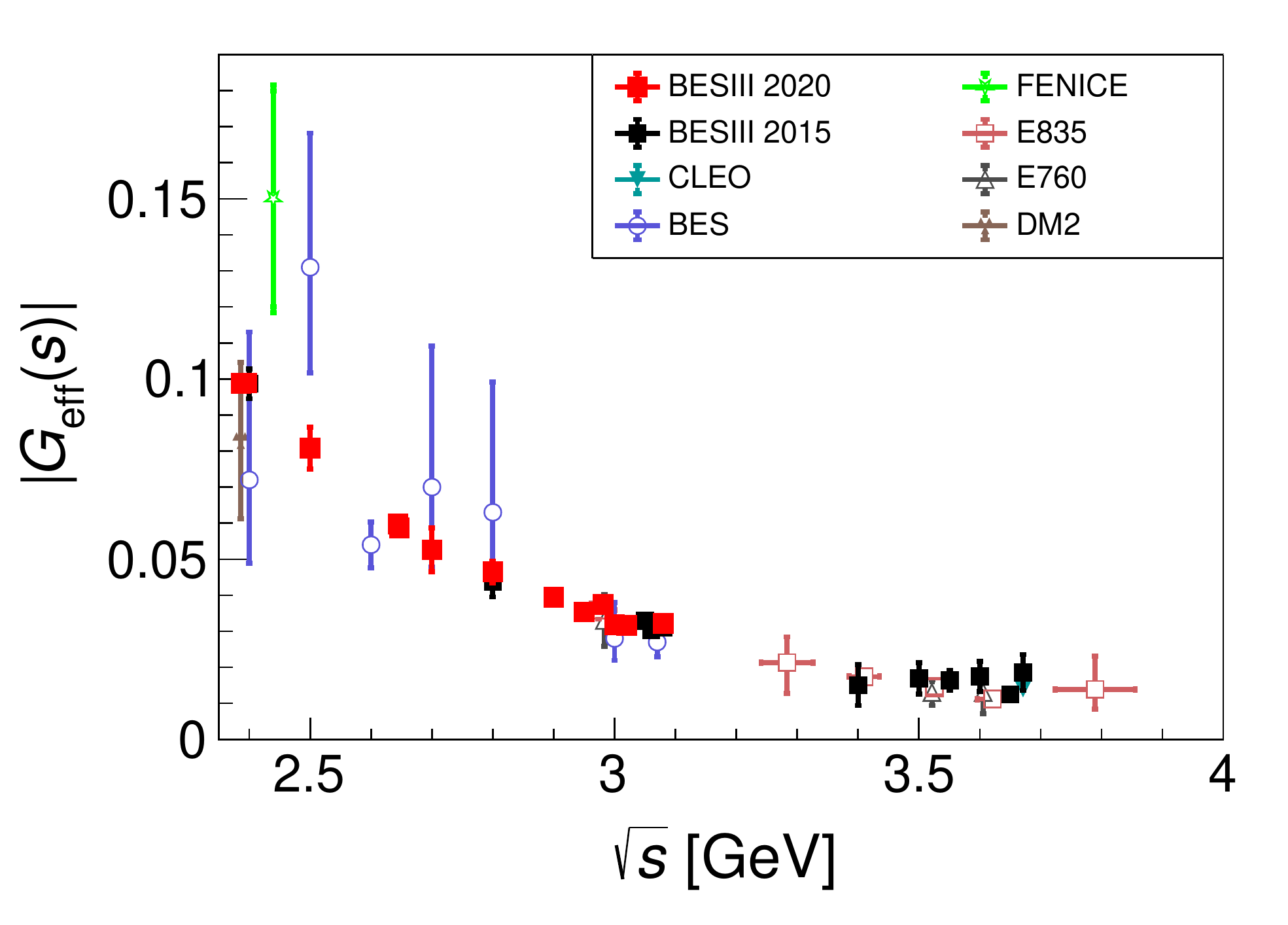}
  \put(0,69){$(b)$}
  \end{overpic}
 }

 \vskip-5.0pt
 \caption{
Comparison of the results for $|G_{\text{eff}}|$ using the energy scan strategy in (a) $2m_{p}<\sqrt{s}<2.35$~GeV,  (b) $2.35<\sqrt{s}<4.00$~GeV.
Shown are all the published measurements from BESIII~\cite{xiaorong,BES3scanLXRC}, CMD3~\cite{cmd3}, CLEO~\cite{introCLEO}, BES~\cite{introBES}, FENICE~\cite{introFENICE}, E835~\cite{introE835}, E760~\cite{introE760}, PS170~\cite{introPS170}, DM2~\cite{introDM2}, DM1~\cite{introDM1}, M.~S.~T~\cite{MSTColl2}, and ADONE~\cite{Adone}.
}
\label{fig_effFF_result1}
\end{center}
\end{figure}
A comparison of $|G_{\text{eff}}|$ of the proton determined by the different experiments is shown in Figure~\ref{fig_effFF_result1}~(a) for c.m.~energies from threshold up to 2.35~GeV, and in Figure~\ref{fig_effFF_result1}~(b) for c.m.~energies between 2.35 and 4.00~GeV.
Within their respective uncertainties, the measurements agree well with each other. 
Measurements close to the threshold, especially by the PS170 experiment, show a steep rise of $|G_{\text{eff}}|$ towards the threshold, while data at higher momentum transfer mostly follows a dipole behavior. 
A deviation from this behavior can only be seen in the most precise measurement recently performed at BESIII, where a periodic behavior was found to be superimposed over the monotonous decrease of $|G_{\text{eff}}|$. 
The origin of this structure is still under debate, with possible sources being rescattering of the forming final state particles or intermediate resonance states before the formation of the proton-antiproton state which enhances the cross section of the process within a certain momentum transfer region. A more detailed discussion of this phenomenon can be found in Section~\ref{sec_GeffPeriodic}.

\subsection{Measurements of the FF ratio $|G_{E}/G_{M}|$ of the proton}
\label{sec_RComp}

Measurements with the scan technique of the individual proton TL EM FFs over a wider momentum transfer range are limited to three published measurements, two of which were performed with the BESIII detector. 
A determination of the ratio of the electric and magnetic FF requires an angular analysis of the data, while a separate measurement of $|G_{E}|$ and $|G_{M}|$ additionally requires precise knowledge of the luminosity of the acquired data. 
For most proton TL EM FF measurements, either this knowledge was missing or the amount of detected events did not allow for a measurement of the differential cross section.

\begin{figure}[htbp]
\begin{center}
\centering
\vskip-0pt
\mbox{
  \begin{overpic}[width=6.6cm,height=5.0cm,angle=0]{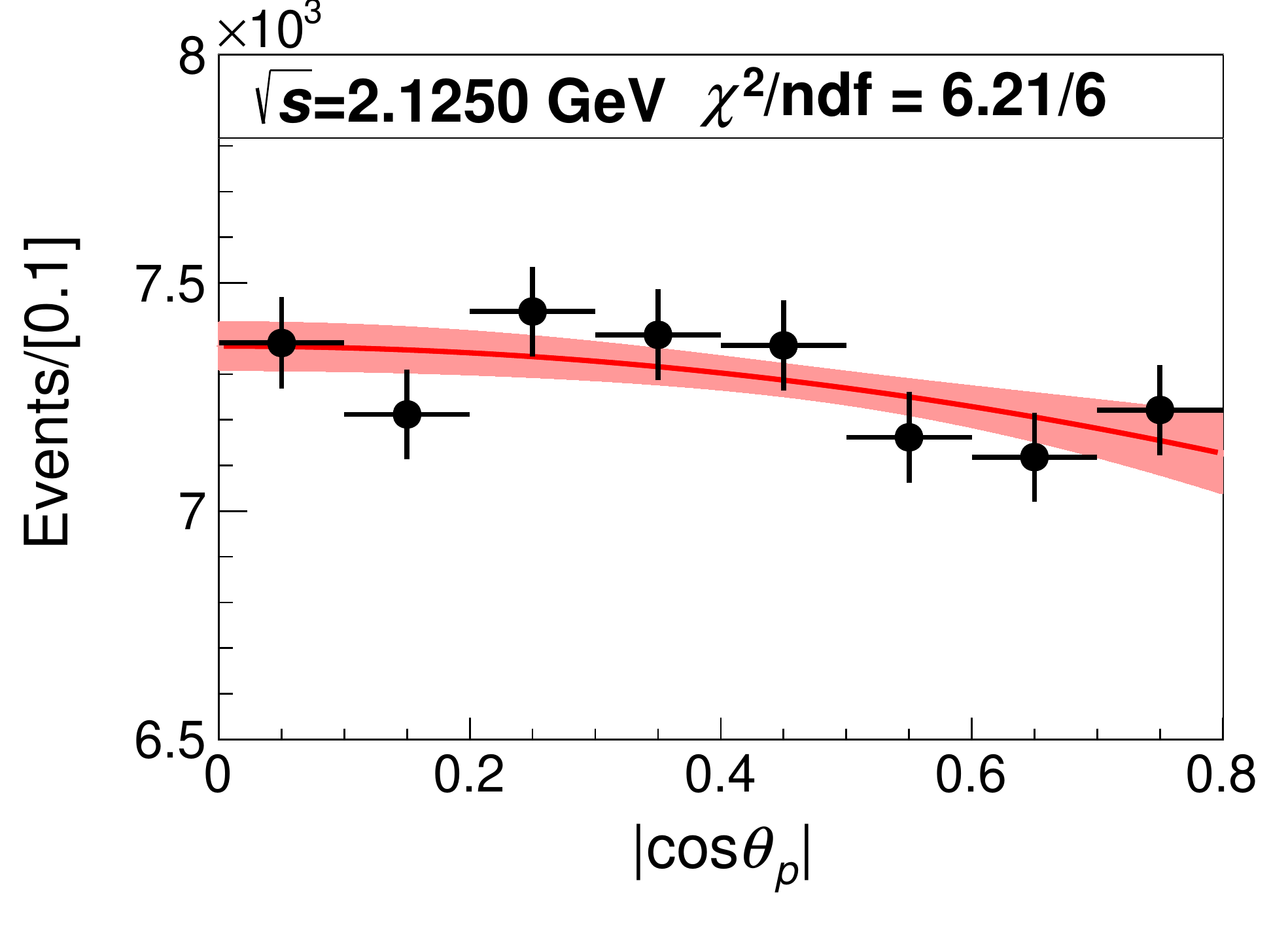}
  \put(0,69){$(a)$}
  \end{overpic}
 \begin{overpic}[width=6.6cm,height=5.0cm,angle=0]{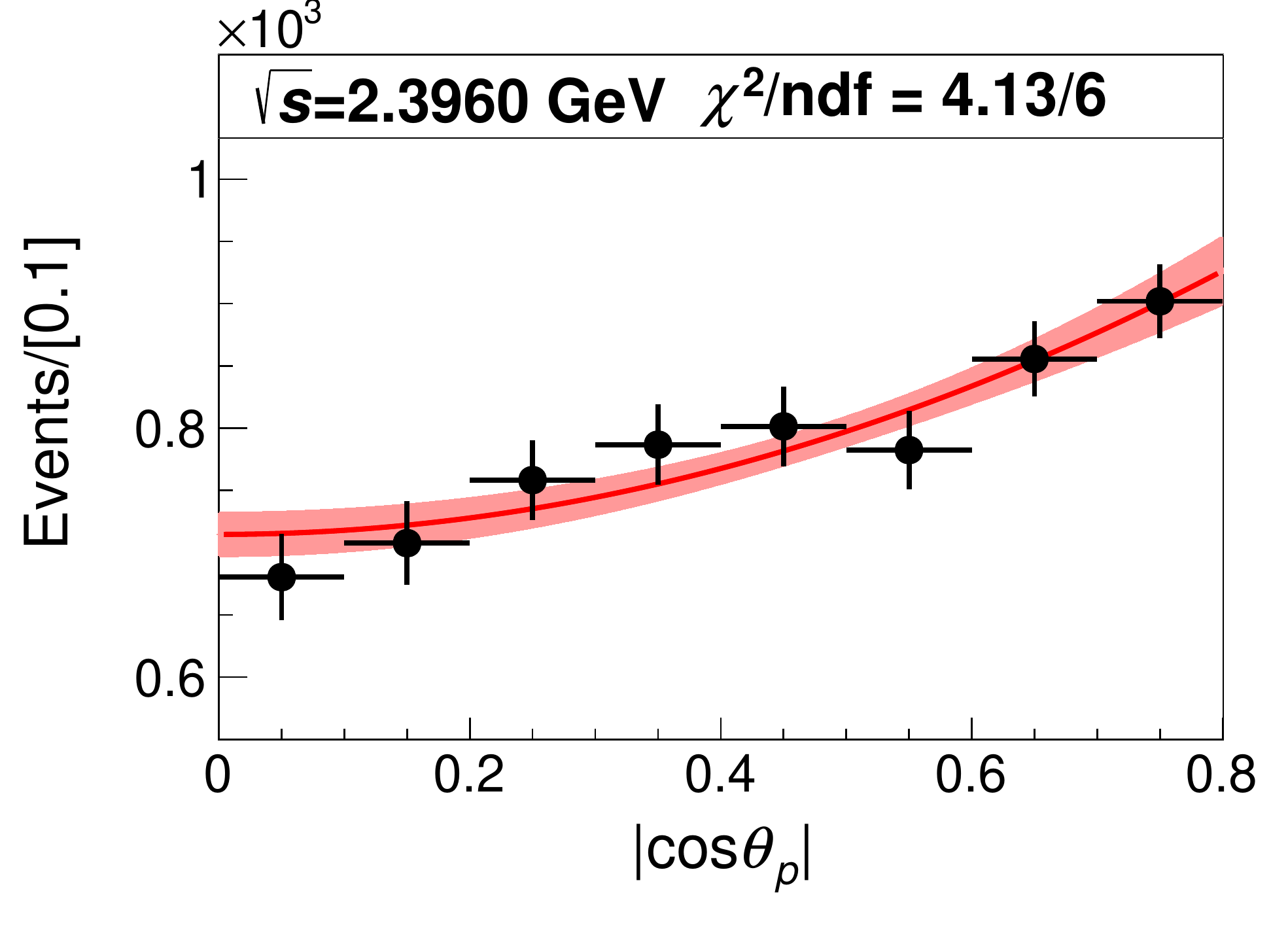}
  \put(0,69){$(b)$}
  \end{overpic}
 }

 \vskip-5.0pt
 \caption{
Fit to the  $|\cos\theta|$ distributions at 2.125~GeV and 2.396~GeV at BESIII~\cite{BES3scanLXRC}, after the application of angular-dependent $\epsilon(1 + \delta)$ factors.
}
\label{fig_cosfit_final}
\end{center}
\end{figure}

In one-photon approximation, the FFs $|G_{E}|$ and $|G_{M}|$, or equivalently their ratio $|G_{E}/G_{M}|$ and $|G_{M}|$ can be determined from a fit to the proton angular distribution for energy points with a sufficiently high number of selected candidates.
The range of the angular analysis in most experiments is limited to $\cos\theta$ from $-0.8$ to $0.8$.
The formula used to fit the proton angular distribution, deduced from Eq.~(\ref{eq_FF3}), can be expressed as
\begin{eqnarray}
\label{eq_fit}
\begin{split}
\frac{dN}{\epsilon\left(1+\delta\right)\times d\cos\theta} = \frac{\mathcal{L}_{\text{int}}\pi\alpha^{2}\beta C}{2s}|G_{M}|^{2}\Bigl[\left(1+\cos^{2}\theta\right)+\frac{4m_p^{2}}{s}\left|\frac{G_{E}}{G_{M}}\right|^{2}\left(1-\cos^{2}\theta\right)\Bigr],
\end{split}
\end{eqnarray}
where $\epsilon(\cos\theta)$ is the angular-dependent efficiency, including both the detector efficiency and the efficiency from the event selection, and $(1+\delta)(\cos\theta)$ is the correction factor for both radiative corrections, such as ISR and final state radiation (FSR), as well as vacuum polarisation effects. Both correction factors are obtained from Monte Carlo (MC) simulation.
After applying the corrections, the $|\cos\theta|$ distribution is fitted with Eq.~(\ref{eq_fit}). 
An example of the angular distribution of outgoing protons including such a fit is shown in Figure~\ref{fig_cosfit_final} at 2.125~GeV and 2.396~GeV from the most recent BESIII measurement~\cite{BES3scanLXRC}.

\begin{figure}[htbp]
\begin{center}
\centering
\vskip-0pt
\mbox{
  \begin{overpic}[width=10.0cm,height=7.5cm,angle=0]{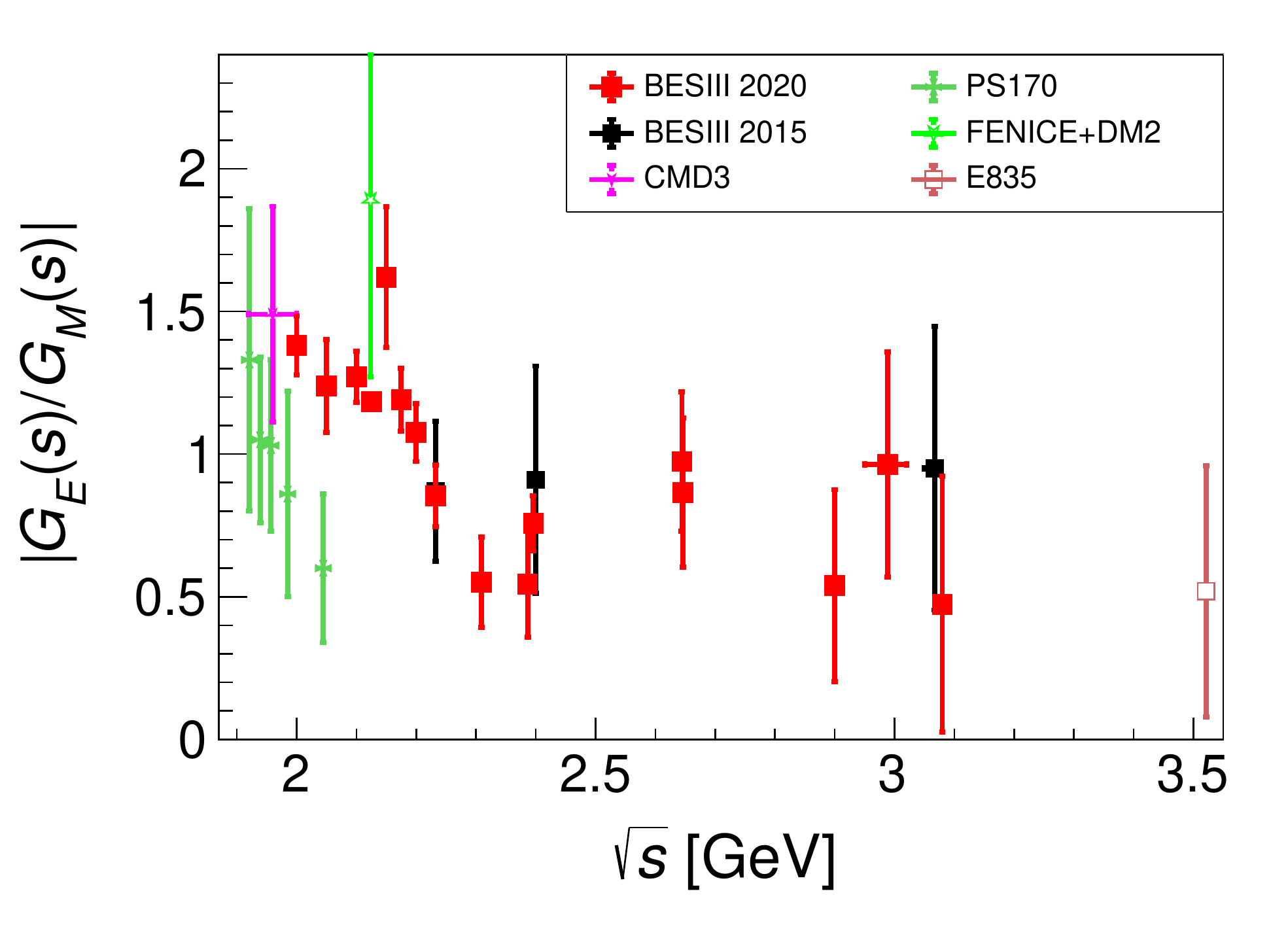}
  \end{overpic}
 }

 \vskip-5.0pt
 \caption{
Comparison of the results for $|G_{E}/G_{M}|$ using the energy scan strategy.
Shown are all the published measurements from BESIII~\cite{xiaorong,BES3scanLXRC}, CMD3~\cite{cmd3}, PS170~\cite{introPS170}, FENICE+DM2~\cite{introFENICE,introDM2,Rinaldo2}, E835~\cite{introE835, Rinaldo2}.
}
\label{fig_Ratio_result}
\end{center}
\end{figure}

The published results for the existing measurements of the $|G_{E}/G_{M}|$ using the scan technique are shown in Figure~\ref{fig_Ratio_result}. 
The most recent BESIII results are the most extensive and precise ones. 
They agree well with the previous BESIII results obtained with smaller statistics. The single measurement point from CMD3 is also in good agreement within its uncertainties.
In contrast, the PS170 results show a systematic trend towards smaller values of $|G_{E}/G_{M}|$, which is not confirmed by the BESIII results.

\subsection{Measurements of the $|G_{E}|$ and $|G_{M}|$ of proton}
\label{sec_FFComp}

\begin{figure}[htbp]
\begin{center}
\centering
\vskip-0pt
\mbox{
  \begin{overpic}[width=6.6cm,height=5.0cm,angle=0]{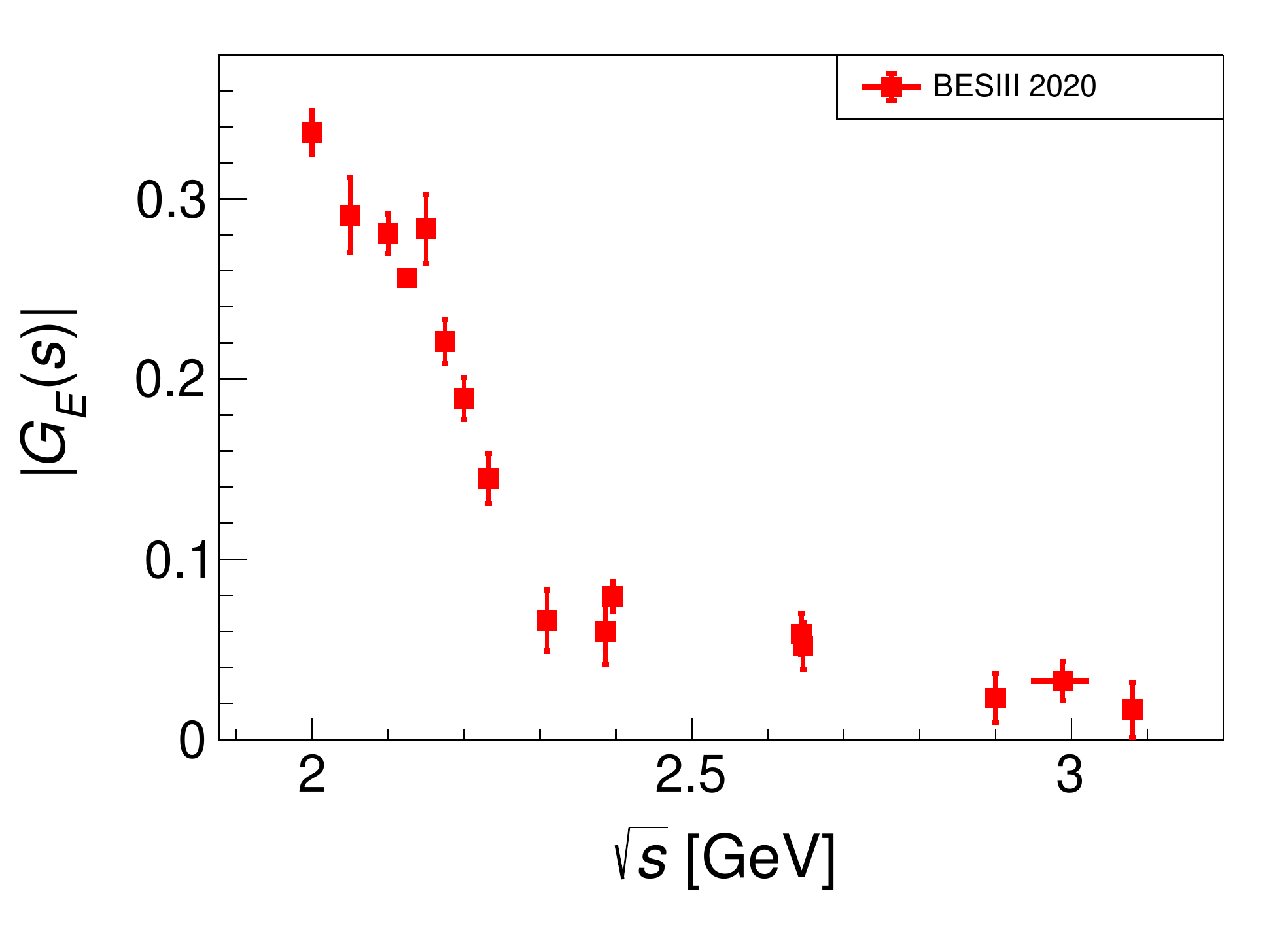}
  \put(0,69){$(a)$}
  \end{overpic}
 \begin{overpic}[width=6.6cm,height=5.0cm,angle=0]{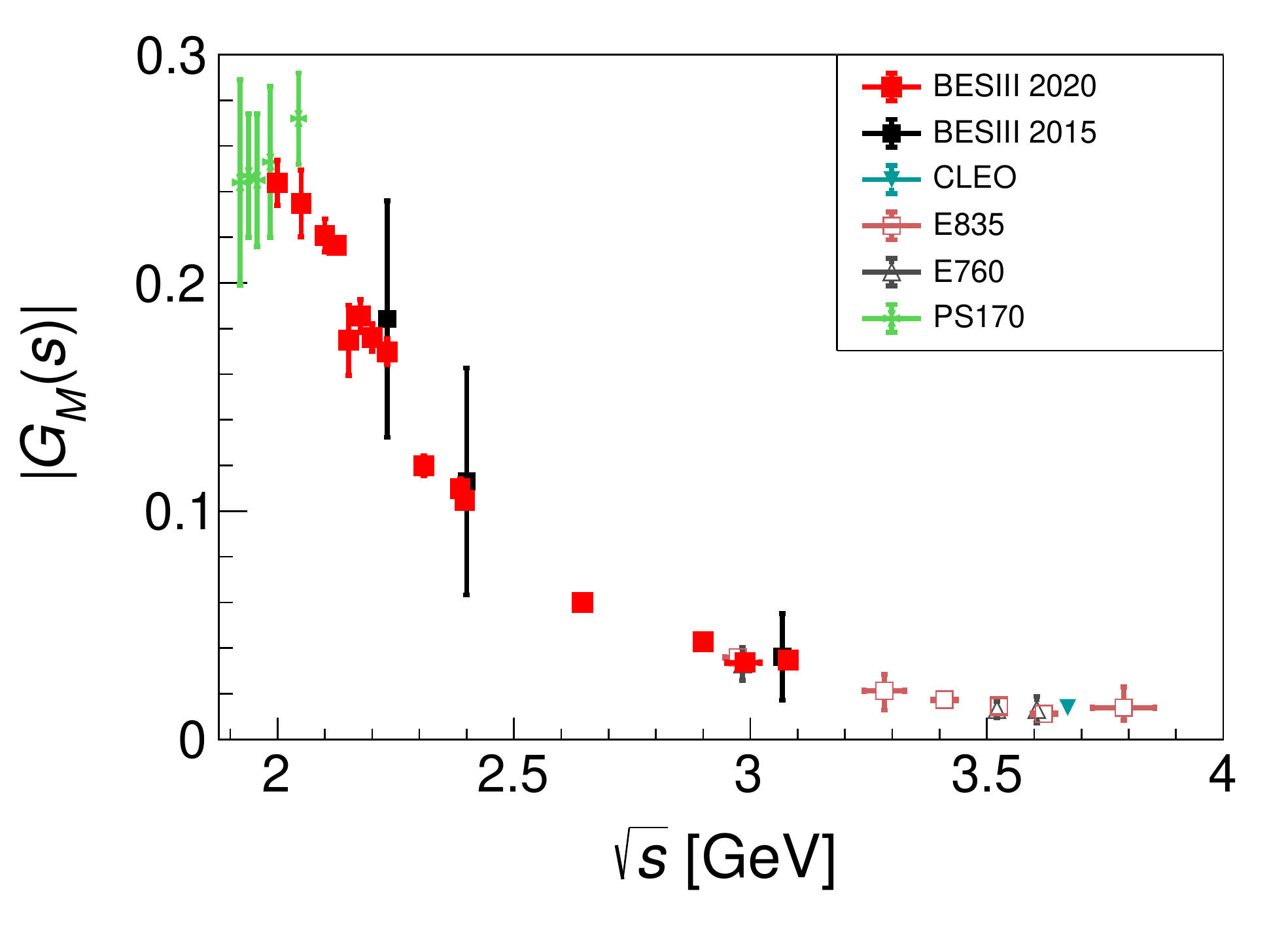}
  \put(0,69){$(b)$}
  \end{overpic}
 }

 \vskip-5.0pt
 \caption{
Comparison of the results for (a) $|G_{M}|$ and (b) $|G_{E}|$ using the energy scan strategy.
Shown are all the published measurements from BESIII~\cite{xiaorong,BES3scanLXRC}, CLEO~\cite{introCLEO}, E835~\cite{introE835}, E760~\cite{introE760}, PS170~\cite{introPS170}.
}
\label{fig_FF_result}
\end{center}
\end{figure}

In the case of $|G_{E}|$, the only published results are from the recent BESIII high luminosity scan measurement. However, both the smaller, previous BESIII scan as well as PS170 have measured $|G_{E}/G_{M}|$ and $|G_{M}|$, which would allow for a calculation of $|G_{E}|$.
The sole direct measurement of $|G_{E}|$ is shown on the left side of Figure~\ref{fig_FF_result}~(a), while the right side (b) shows a comparison of the results for $|G_{M}|$. 

For $|G_{M}|$, the available measurements are all in agreement, albeit the BESIII measurement from 2020 is by far the most precise. It should be noted that the three measurements at high momentum transfer by CLEO, E780, and E835 were performed under the assumption of $|G_{E}|=0$ instead of extraction through angular analysis. Therefore, they are more of an $|G_{\text{eff}}|$, albeit the assumption may be justified due to the strong suppression of $|G_{E}|$ at higher energies.

Both the $|G_{E}/G_{M}|$ as well as the individual FF measurement, especially that of $|G_{E}|$, seem to show a periodic structure on top of a dipole like, monotonously falling behavior. 
This structure is similar to the one already observed in $|G_{\text{eff}}|$, however, in the case of the individual FFs it is only visible for the high precision fine scan data from 2020. 
First approaches to describe these features have been made from the theoretical side with mostly phenomenological approaches. However, more investigations especially concerning the origin of these structures, are necessary. 
From the experimental side, independent confirmation of the BESIII measurement of $|G_{E}/G_{M}|$, $|G_{E}|$ and $|G_{M}|$ is needed to confirm this behavior. An attempt at a phenomenological description of these structures can be found in Section~\ref{sec_TheoRatio}.

\section{Theoretical approaches to the description of Timelike Electromagnetic Proton Form Factors}
\subsection{Test of perturbative QCD and low energy effective theory for QCD}
\label{pQCD_low-enegy-eff-QCD}
The recently obtained high precision results in the TL region provide information to improve our understanding of the proton inner structure and test theoretical models that depend on perturbative QCD (pQCD) and low energy effective theory for QCD (low-enegy-eff-QCD). The low energy region is the regime of low-enegy-eff-QCD due to the growing of the running QCD coupling constant and the associated confinement of quarks and gluons. 
The Energy region between 2.00 and 3.08~GeV connects the low-enegy-eff-QCD and the pQCD regime. 
Therefore, Ref.~\cite{BES3scanLXRC} suggests a fit function for the energy dependence of $\sigma_{p\bar{p}}$ within this energy region, consisting of a low-enegy-eff-QCD and a pQCD part:
\begin{eqnarray}
\label{eq_fitXsec_2}
\sigma_{p\bar{p}} (s)=\left\{
\begin{aligned}
 \begin{split}
\sigma_{\text{eff-QCD}}(s)&=\frac{e^{a_{c}}\pi^{2}\alpha^{3}}{s\left[1-e^{-\frac{\pi\alpha_{s}(s)}{\beta(s)}}\right]\left[1+\left(\frac{\sqrt{s}-2m_{p}c^{2}}{a_{0}}\right)^{a_{1}}\right]},~\sqrt{s}\leq2.3094~{\text{GeV}},\\
 \end{split}\\
\begin{split}
\sigma_{\text{pQCD}}(s)&=\frac{2\pi\alpha^{2}\beta(s)C\left[2+\left(\frac{2m_{p}c^{2}}{\sqrt{s}}\right)^{2}\right]e^{2a_{2}}}{3s^{5}\left[4\ln^{2}\left(\frac{\sqrt{s}}{a_{0}}\right)+\pi^{2}\right]^{2}},~\sqrt{s}>2.3094~{\text{GeV}},
\end{split}
\end{aligned}
\right.
\end{eqnarray}
with the strong coupling constant $\alpha_{s}(s)$ and the fine-structure constant $\alpha$. For the strong coupling constant, the following parametrization is used:
\begin{eqnarray}
\label{alphabeta}
\alpha_{s}(s) &=\left[\frac{1}{\alpha_{s}\left(m_{Z}^{2}c^{4}\right)}+\frac{25}{12\pi}\ln\left(\frac{s}{m_{Z}^{2}c^{4}}\right)\right]^{-1},
\end{eqnarray}
with the mass of the $Z$ boson $m_{Z} = 91.1876$~GeV/$c^{2}$ and the strong coupling constant at the $Z$ pole $\alpha_{s}(m_{Z}^{2}c^{4})=0.11856$.
Instead of introducing the Coulomb enhancement factor, the suggested fit function in Ref.~\cite{BES3scanLXRC} (Eq.~\ref{eq_fitXsec_2}) uses an alternative approach close to the $p\bar{p}$ threshold by considering gluon exchange between the $B\bar{B}$ pair. In addition, the first part of Eq.~(\ref{eq_fitXsec_2}) describes the behavior near threshold by considering strong interaction effects from the low-enegy-eff-QCD for low the energy region near the $p\bar{p}$ threshold~\cite{Solovtsova,Rinaldo3}.
The second part of the formula describes the pQCD behavior of the cross section, computed in leading order for the continuum region at high momentum transfer~\cite{Bianconi}.
 
\begin{figure}[htbp]
\begin{center}
\centering
\vskip-0pt
 \mbox{
 \begin{overpic}[width=10.0cm,height=7.5cm,angle=0]{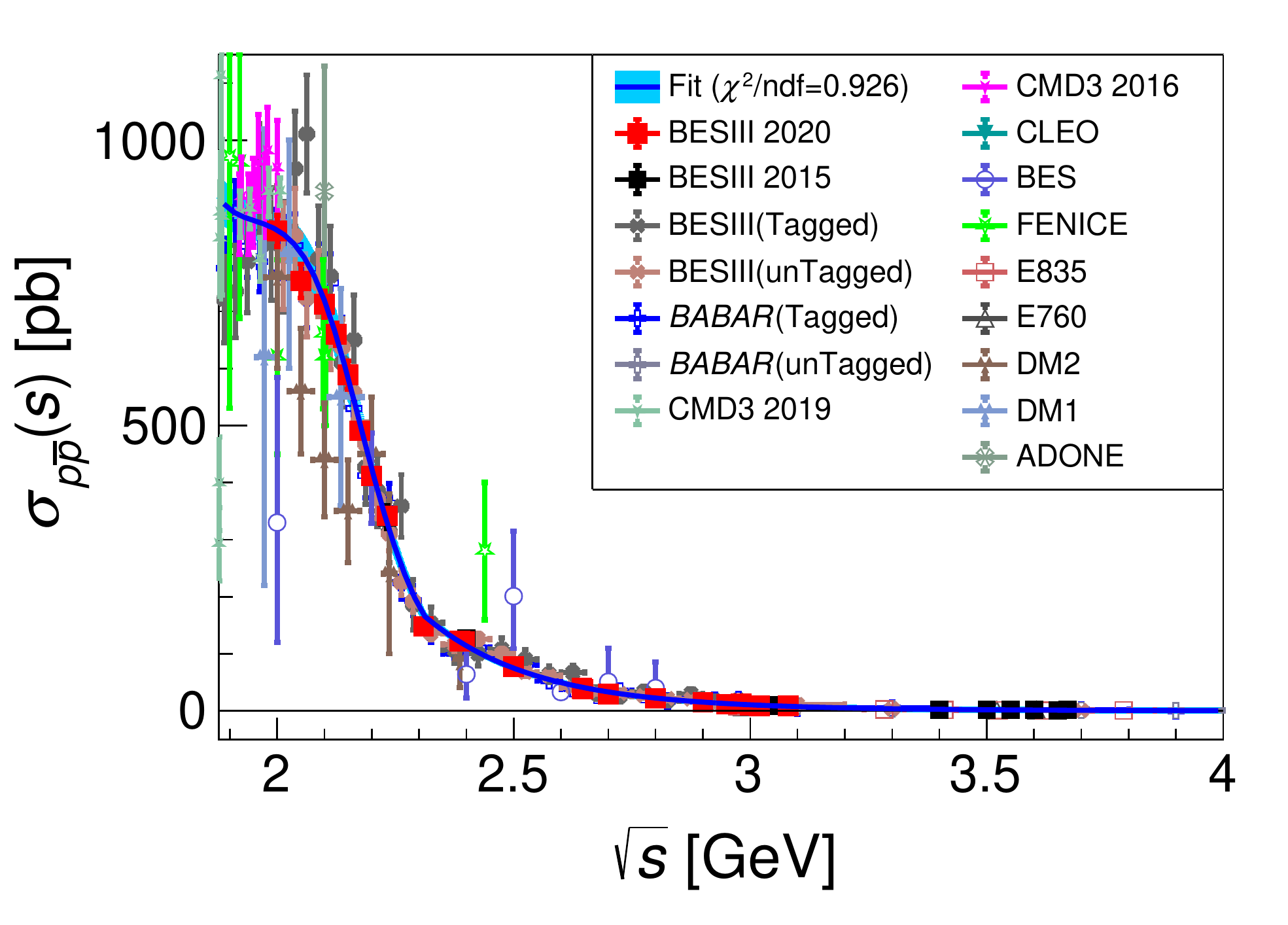}
  \end{overpic}
 }
 \vskip-0.0pt
 \caption{Summary for the $e^{+}e^{-}\xleftrightarrow{}p\bar{p}$ cross section and a fit to the data (blue solid line and band) according to Eq.~(\ref{eq_fitXsec_2}).
The shown data are the published measurements from BESIII~\cite{xiaorong,BES3untagged,BES3scanLXRC,BES3tagged}, {\textsl{BABAR}}~\cite{introBABAR}, CMD3~\cite{cmd3}, CLEO~\cite{introCLEO}, BES~\cite{introBES}, FENICE~\cite{introFENICE}, E835~\cite{introE835}, E760~\cite{introE760}, DM2~\cite{introDM2}, DM1~\cite{introDM1}, and ADONE~\cite{Adone}.
In the fit, $\chi^{2}$ is defined as $\chi^{2}=\sum_{i}[f(x_{i})-y_{i}]^{2}/{\text{err}}_{i}^{2}$, where $\text{err}_{i}$ is the error of the measured results including statistical and correlated systematic uncertainties, $f$ is the fit function, and n.d.f. is the number of degrees of freedom. 
}
\label{fig_Xsec_low-enegy-eff-QCD}
\end{center}
\end{figure}
In conclusion, the low-enegy-eff-QCD based description is suitable for the lower energy behavior below 2.3094 GeV, while the pQCD based approach provides a good description for the cross section data at higher energy, which is illustrated in Figure~\ref{fig_Xsec_low-enegy-eff-QCD}.
The results and meaning of the fit parameters are as follows: $a_{0}=332\pm3$~MeV is the overall QCD parameter,
$a_{1}=4.78\pm0.13$ is the $\sigma_{p\bar{p}}(s)$ power-law dependence, which is related to the number of valence quarks, and $a_{2}=4.39\pm0.01$ is a normalization constant.
Finally, $a_{c}$ makes $\sigma_{\text{low-enegy-eff-QCD}}({\text{2.3094~GeV}})=\sigma_{\text{pQCD}}({\text{2.3094~GeV}})$, and does not attend the global fit.

\subsection{Possible Resonance Structures in the $p\bar{p}$ system}
\label{sec_Xsec_Res}
A description of the cross section of $e^{+}e^{-}\xleftrightarrow{} p\bar{p}$ according to a pQCD and low-enegy-eff-QCD model has been introduced in Section~\ref{pQCD_low-enegy-eff-QCD}.
The slight ridges and bump structures occurring here imply that the model in Section~\ref{pQCD_low-enegy-eff-QCD} and its description of the low-enegy-eff-QCD and pQCD regime are not perfect.
Various approaches have been proposed to describe these deviations, including resonance structures~\cite{Lorenz} and periodic interference structures~\cite{Bianconi}.
In the first instance, possible resonance structures are considered as convex modifications to the overall concave function describing $|G_{\text{eff}}|$ at invariant masses of around 2.00–2.25 GeV. 
The approach will be introduced in the following, and the fit parameters are extracted with improved precision by including the new high luminosity data from BESIII. The line shape of $|G_{\text{eff}}|$ is fitted using a coherent sum of a Breit-Wigner function and a nonresonant term~\cite{Lorenz}:
\begin{eqnarray}
\label{eq_Xsec_Res1}
\sigma_{e^{+}e^{-}\xleftrightarrow{} p\bar{p}}(s)=\frac{4\pi\alpha^{2}\beta C}{3s}\left(1+\frac{2m_{p}^{2}}{s}\right)|G_{\text{eff}}^{\text{n.r.}}+A_{p\bar{p}}^{\text{Res(2150)}}|^{2}.
\end{eqnarray}
The first term $G_{\text{eff}}^{\text{n.r.}}$ is the nonresonant contribution,
\begin{eqnarray}
\label{eq_FFnonres}
G_{\text{eff}}^{\text{n.r.}}=\sqrt{\frac{3s\sigma_{p\bar{p}}(s)}{4\pi\alpha^{2}\beta C(1+\frac{2m_{p}^{2}}{s})}},
\end{eqnarray}
where $\sigma_{p\bar{p}}(s)$ is from Eq.~\ref{eq_fitXsec_2}.
The second term $A_{p\bar{p}}^{\text{Res(2150)}}$ is the Breit-Wigner amplitude:
\begin{eqnarray}
\label{eq_Ares1}
A_{p\bar{p}}^{\text{Res(2150)}}=\frac{e^{a_{3}+i\Psi_{\text{Res(2150)}}}}{m_{\text{Res}(2150)}^{2}-s-i\Gamma_{\text{Res}(2150)}m_{\text{Res}(2150)}},
\end{eqnarray}
with normalization factor $a_{3}$, mass $m_{\text{Res}(2150)}$, width $\Gamma_{\text{Res}(2150)}$ and relative phase angle to the nonresonant component $\Psi_{\text{Res(2150)}}$.
The results of the fit parameters are as follows: $a_{0}=387_{-22}^{+21}$~MeV, $a_{1}=7.43_{-1.12}^{+1.53}$, $a_{2}=4.33_{-0.08}^{+0.06}$, $a_{3}=7.73_{-0.18}^{+0.12}$,  $m_{\text{Res}(2150)}=2150_{-12}^{+10}$~MeV/$c^{2}$, $\Gamma_{\text{Res}(2150)}=359_{-23}^{+30}$~MeV and  $\Psi_{\text{Res(2150)}}=-1.17_{-0.13}^{+0.11}$.
This structure at around 2.15~GeV can be attributed, e.g., to the $\rho(2150)$ resonance~\cite{Rho2150}.
\begin{figure}[htbp]
\begin{center}
\centering
\vskip-0pt
 \mbox{
 \begin{overpic}[width=6.6cm,height=5.0cm,angle=0]{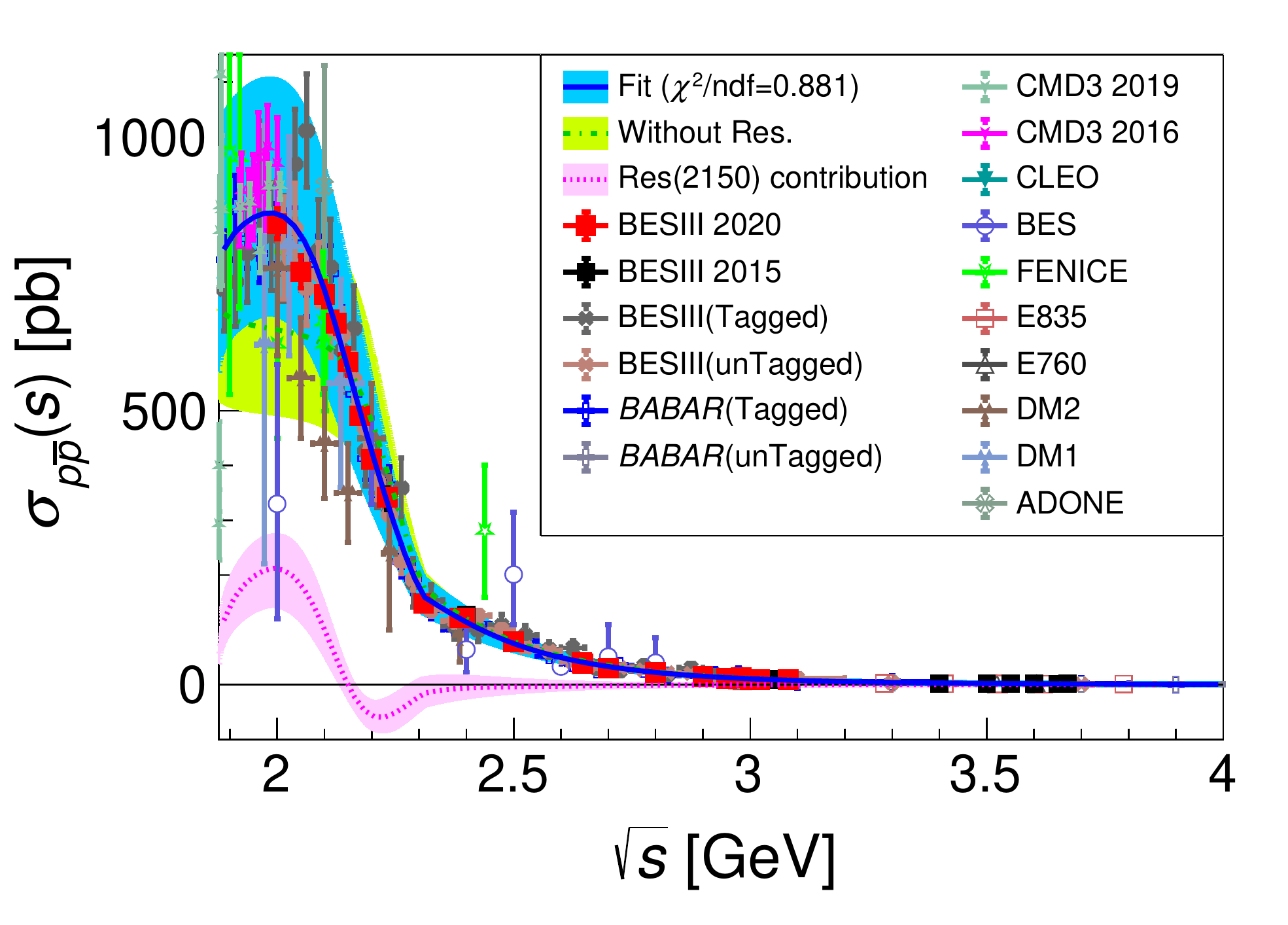}
  \put(0,69){$(a)$}
  \end{overpic}
  \begin{overpic}[width=6.6cm,height=5.0cm,angle=0]{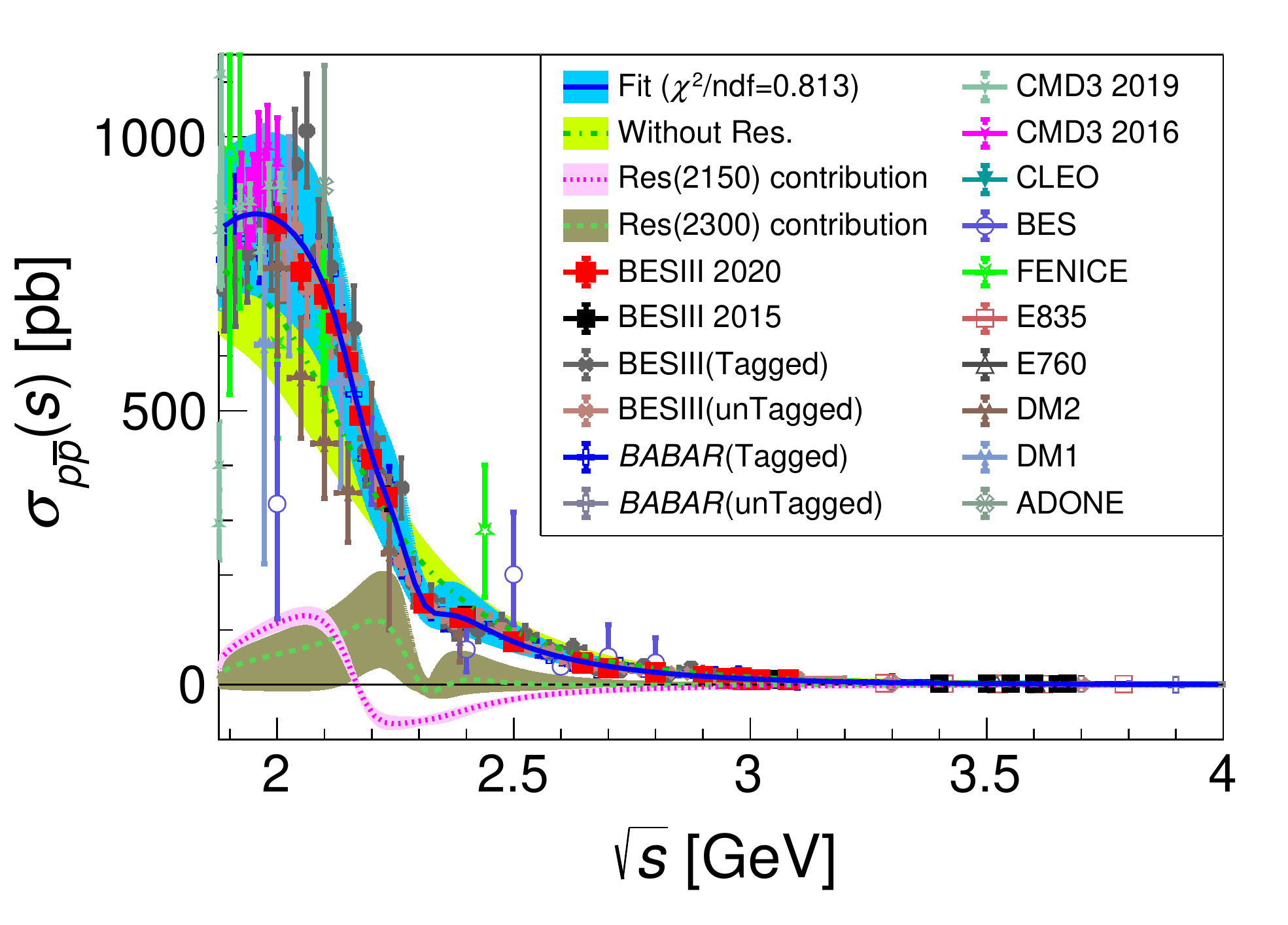}
  \put(0,69){$(b)$}
  \end{overpic}
 }
 \vskip-0.0pt
 \caption{Summary for the $e^{+}e^{-}\xleftrightarrow{}p\bar{p}$ cross section and a fit to the data (blue solid line and band) according to (a) Eq.~(\ref{eq_Xsec_Res1}) and (b) Eq.~(\ref{eq_Xsec_Res2}), with the low-enegy-eff-QCD and pQCD model (green dash-dotted curve and band), and contributions from possible resonances at around 2.15~GeV (magenta dotted curve and band) and around 2.3~GeV (teal dashed curve and band).
The data shown are the published measurements from BESIII~\cite{xiaorong,BES3untagged,BES3scanLXRC,BES3tagged}, {\textsl{BABAR}}~\cite{introBABAR}, CMD3~\cite{cmd3}, CLEO~\cite{introCLEO}, BES~\cite{introBES}, FENICE~\cite{introFENICE}, E835~\cite{introE835}, E760~\cite{introE760}, DM2~\cite{introDM2}, DM1~\cite{introDM1}, and ADONE~\cite{Adone}.
The definition of $\chi^{2}$ is the same as in Figure~\ref{fig_Xsec_low-enegy-eff-QCD}. 
}
\label{fig_Xsec_res1}
\end{center}
\end{figure}

An additional resonance at around 2.3 GeV can also be considered in the fit~\cite{Lorenz}:
\begin{eqnarray}
\label{eq_Xsec_Res2}
\sigma_{e^{+}e^{-}\xleftrightarrow{} p\bar{p}}(s)=\frac{4\pi\alpha^{2}\beta C}{3s}\left(1+\frac{2m_{p}^{2}}{s}\right)|G_{\text{eff}}^{\text{n.r.}}+A_{p\bar{p}}^{\text{Res(2150)}}+A_{p\bar{p}}^{\text{Res(2300)}}|^{2},
\end{eqnarray}
where $A_{p\bar{p}}^{\text{Res(2300)}}$ is the Breit-Wigner amplitude:
\begin{eqnarray}
\label{eq_Ares2}
A_{p\bar{p}}^{\text{Res(2300)}}=\frac{e^{a_{4}+i\Psi_{\text{Res(2300)}}}}{m_{\text{Res}(2300)}^{2}-s-i\Gamma_{\text{Res}(2300)}m_{\text{Res}(2300)}},
\end{eqnarray}
with normalization factor $a_{4}$, mass $m_{\text{Res}(2300)}$, width $\Gamma_{\text{Res}(2300)}$ and relative phase angle to the nonresonant component $\Psi_{\text{Res(2300)}}$.
The results of the fit parameters are as follows: $a_{0}=334_{-24}^{+42}$~MeV, $a_{1}=2.66_{-0.65}^{+0.77}$, $a_{2}=4.52_{-0.06}^{+0.04}$, $a_{3}=6.03_{-0.91}^{+0.86}$,  $m_{\text{Res}(2150)}=2169_{-16}^{+26}$~MeV/$c^{2}$, $\Gamma_{\text{Res}(2150)}=187_{-85}^{+60}$~MeV, $\Psi_{\text{Res(2150)}}=0.46_{-0.59}^{+0.51}$, $a_{4}=7.00_{-0.38}^{+0.23}$,  $m_{\text{Res}(2300)}=2302_{-14}^{+16}$~MeV/$c^{2}$, $\Gamma_{\text{Res}(2300)}=188_{-51}^{+34}$~MeV and  $\Psi_{\text{Res(2300)}}=-1.36_{-0.27}^{+0.13}$.
The significance of the two possible resonance structures is determined to be 2.8$\sigma$.

The corresponding work states that this is not a rigorous analysis since these resonance structures overlap and are not separable from the background. 
Nevertheless, the addition of two resonances peaking at around 2.15 and 2.30~GeV significantly improves the description of the data~\cite{Lorenz}.

\subsection{Periodic Interference Structures in the TL Proton FF}
\label{sec_GeffPeriodic}
The second mentioned approach to describe the periodic structures in $|G_{\text{eff}}|$ is presented in this section. The corresponding variables of the fit are determined with improved precision with the new high luminosity data from BESIII.

\begin{figure}[htbp]
\begin{center}
\centering
\vskip-0pt
 \mbox{
 \begin{overpic}[width=6.6cm,height=5.0cm,angle=0]{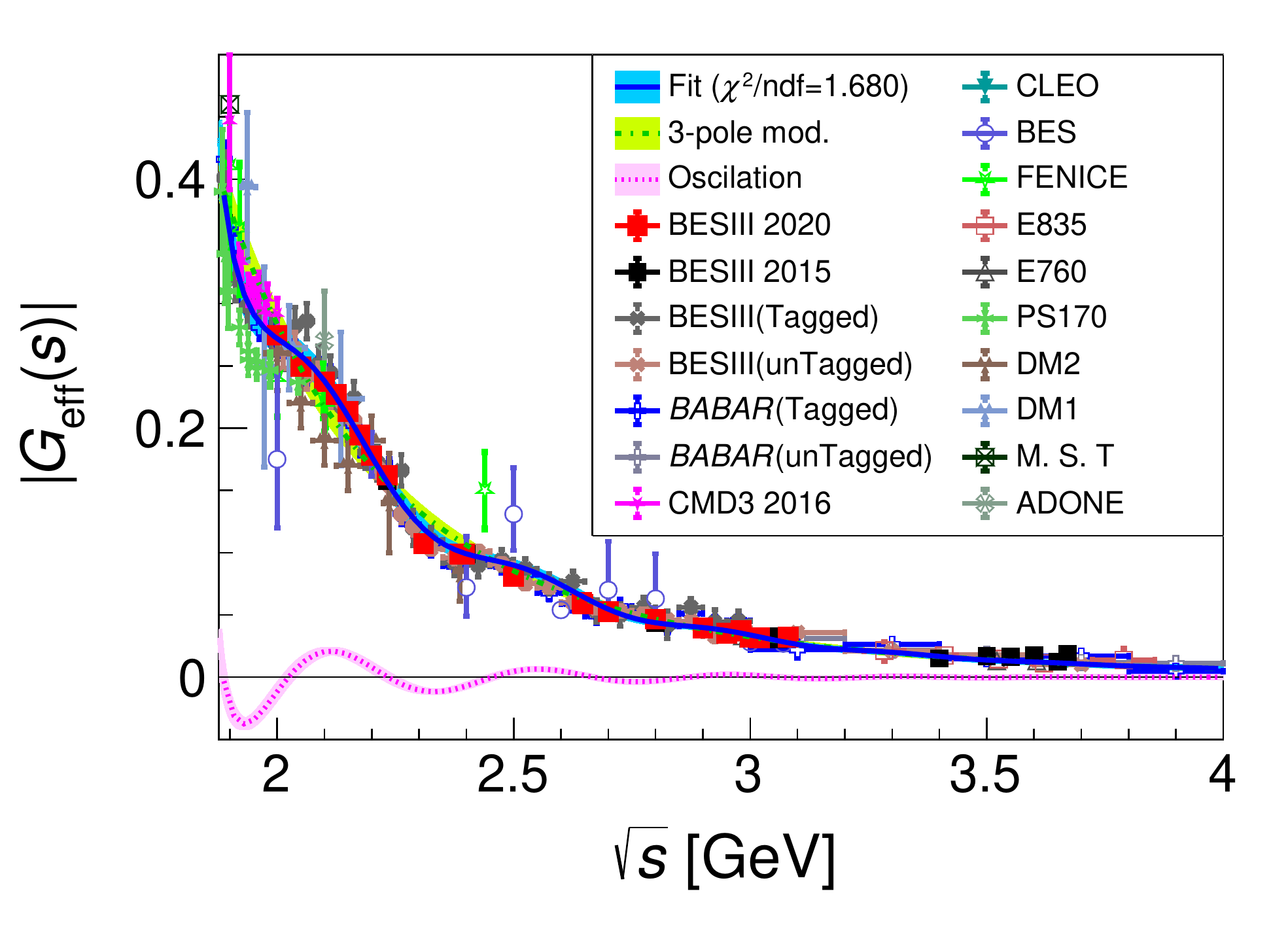}
  \put(0,69){$(a)$}
  \end{overpic}
 \begin{overpic}[width=6.6cm,height=5.0cm,angle=0]{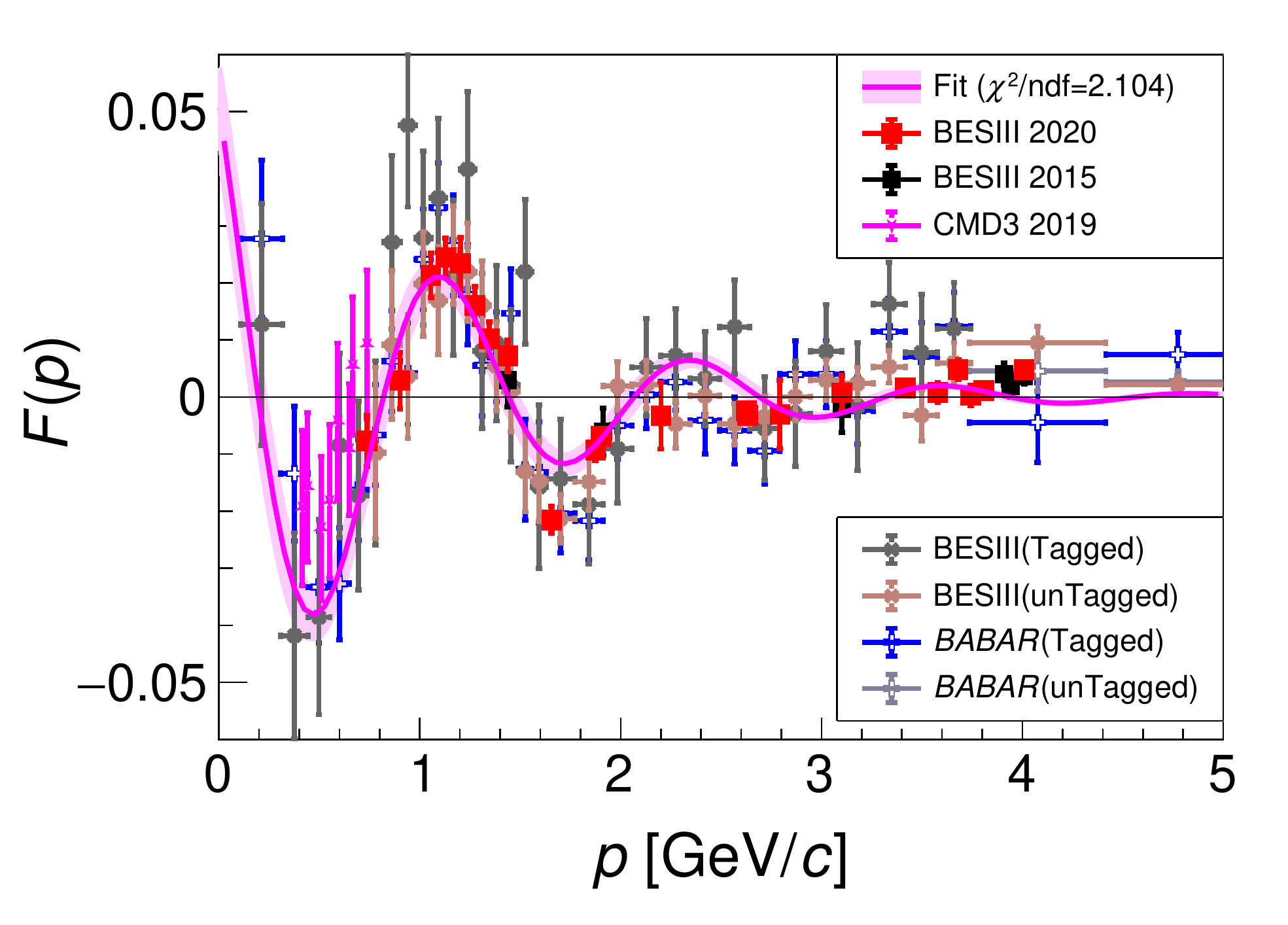}
  \put(0,69){$(b)$}
  \end{overpic} 
  }
 \vskip-0.0pt
 \caption{Summary for (a) the effective FF of the proton $|G_{\text{eff}}|$ and a fit to the data (blue solid line and band) according to Eq.~(\ref{3p-osc}) suggested in Ref.~\cite{Bianconi,Bianconi3} with the 3-pole (green dash-dotted curve and band) and oscillation (magenta dotted curve and band) contributions;
(b) residuals of the proton $|G_{\text{eff}}|$, after subtraction of the smooth function described by Eq.~(\ref{effFFfit}), as a function of the relative momentum $p(s)=\sqrt{s(\frac{s}{4m_{p}^{2}}-1)}$ with a fit of the oscillation according to Eq.~(\ref{osc}) (magenta solid curve and band).
The shown data are the published measurements from BESIII~\cite{xiaorong,BES3untagged,BES3scanLXRC,BES3tagged}, {\textsl{BABAR}}~\cite{introBABAR}, CMD3~\cite{cmd3}, CLEO~\cite{introCLEO}, BES~\cite{introBES}, FENICE~\cite{introFENICE}, E835~\cite{introE835}, E760~\cite{introE760}, PS170~\cite{introPS170}, DM2~\cite{introDM2}, DM1~\cite{introDM1}, M.~S.~T~\cite{MSTColl2}, and ADONE~\cite{Adone}.
The definition of $\chi^{2}$ is the same as in Figure~\ref{fig_Xsec_low-enegy-eff-QCD}.
}
\label{fig_Geff_fit}
\end{center}
\end{figure}

The data of the timelike $|G_{\text{eff}}|$ are best reproduced by the 3-pole model function proposed in Ref.~\cite{Gustafsson},
\begin{eqnarray}
\label{effFFfit}
F_{\text{3p}}(s)=\frac{\mathcal{A}}{(1+\frac{s}{m_{a}^{2}c^{4}})[1-\frac{s}{0.71~\text{GeV}^{2}}]^{2}}.
\end{eqnarray}
A global fit to all available results for $|G_{\text{eff}}|$ using the above equation has been performed and is shown as the green dash-dotted curve and band in Figure~\ref{fig_Geff_fit}~(a).
The residuals between the fit function and the experimental data indicate a periodic structure when they are represented as a function of the relative momentum $p(s)=\sqrt{s(\frac{s}{4m_{p}^{2}}-1)}$ of the $p\bar{p}$ pair~\cite{Bianconi2}.
Phenomenologically, the structure can be described by an oscillating function suggested in Ref~\cite{Bianconi2}: 
\begin{eqnarray}
\label{osc}
F_{\text{osc}}[p(s)]=b^{{\text{osc}}}_{0}e^{-b^{{\text{osc}}}_{1}p(s)}\cos[b^{{\text{osc}}}_{2}p(s)+b^{{\text{osc}}}_{3}].
\end{eqnarray}
The corresponding fit to the residuals is shown as the magenta dotted curve and band in Figure~\ref{fig_Geff_fit}~(a) and the magenta solid curve and band in Figure~\ref{fig_Geff_fit}~(b).
Additionally, a global fit to $|G_{\text{eff}}|$ using the sum of two contributions~\cite{Bianconi, Bianconi3} has been performed:
\begin{eqnarray}
\label{3p-osc}
|G_{\text{eff}}(s)|=F_{\text{3p}}(s)+F_{\text{osc}}[p(s)].
\end{eqnarray}
Here, $\mathcal{A}=8.79_{-0.21}^{+0.22}$, $m_{a}^{2}=9.04_{-0.58}^{+0.62}$~GeV$^{2}/c^{4}$, $b^{{\text{osc}}}_{0}=0.06 \pm 0.01$, $b^{{\text{osc}}}_{1}=0.93 \pm 0.07$~GeV/$c^{-1}$, $b^{{\text{osc}}}_{2}=5.02_{-0.09}^{+0.10}$~GeV/$c^{-1}$ and $b^{{\text{osc}}}_{3}=0.61_{-0.16}^{+0.15}$ are obtained from our fit, which is illustrated as the blue solid line and band in Figure~\ref{fig_Geff_fit}~(a). 

\begin{figure}[htbp]
\begin{center}
\centering
\vskip-0pt
 \mbox{
 \begin{overpic}[width=10.0cm,height=7.5cm,angle=0]{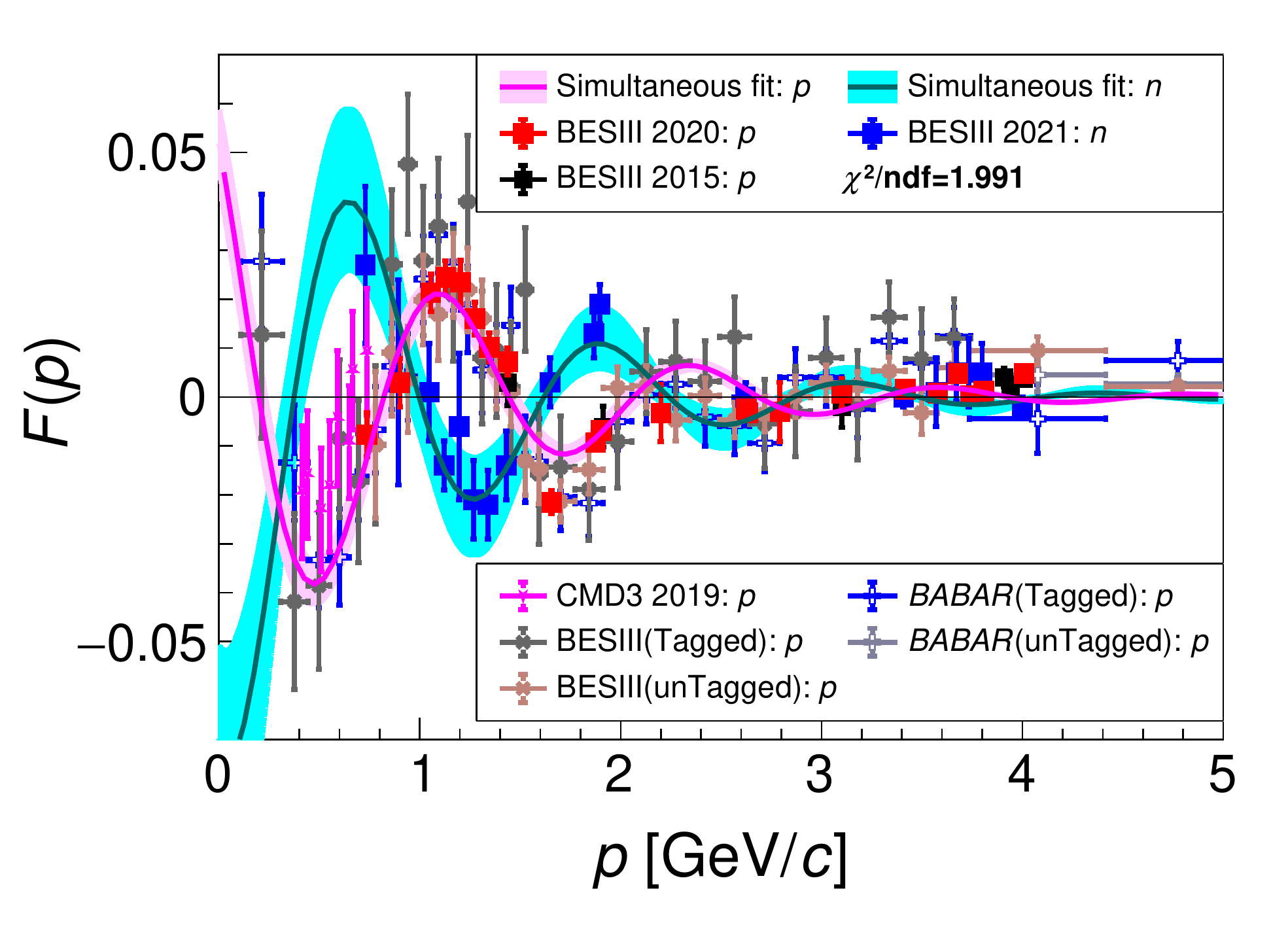}
  \end{overpic}
 }
 \vskip-0.0pt
 \caption{Summary for residuals of the proton and neutron $|G_{\text{eff}}|$, after subtraction of the smooth function described by Eq.~(\ref{effFFfit}), as a function of the relative momentum $p(s)=\sqrt{s(\frac{s}{4m_{p}^{2}}-1)}$ with a fit of the oscillation according to Eq.~(\ref{osc}) (magenta and cyan solid curve and band for proton and neutron, respectively).
The shown data are the published measurements from BESIII (proton~\cite{xiaorong,BES3untagged,BES3scanLXRC,BES3tagged} and neutron~\cite{neutron}), {\textsl{BABAR}}~\cite{introBABAR} and CMD3~\cite{cmd3}.
The definition of $\chi^{2}$ is the same as in Figure~\ref{fig_Xsec_low-enegy-eff-QCD}.
}
\label{fig_Geffpn_fit}
\end{center}
\end{figure}

Periodic interference structures of $|G_{\text{eff}}|$ for the proton manifest as a deviation from a modified dipole behavior.
A similar oscillation with a comparable frequency is observed for the neutron, albeit with a significant phase difference, as illustrated in Figure~\ref{fig_Geffpn_fit}. 
The fit yields $b^{{\text{osc}}}_{2}=5.05_{-0.09}^{+0.10}$~GeV/$c^{-1}$, the phase of proton, $b^{{\text{osc}}}_{3(p)}=0.57_{-0.16}^{+0.14}$, $b^{{\text{osc}}}_{3(n)}=2.82_{-0.21}^{+0.22}$, and a phase difference of $\Delta b^{{\text{osc}}}_{3}=|b^{{\text{osc}}}_{3(p)}-b^{{\text{osc}}}_{3(n)}|=(129_{-10}^{+11})^{\circ}$
The result indicates unexplored intrinsic dynamics that lead to almost orthogonal oscillations.

\subsection{Phenomenological analysis of $|G_{E}/G_{M}|$}
\label{sec_TheoRatio}

\begin{figure}[htbp]
\begin{center}
\centering
\vskip-0pt
 \mbox{
 \begin{overpic}[width=10.0cm,height=7.5cm,angle=0]{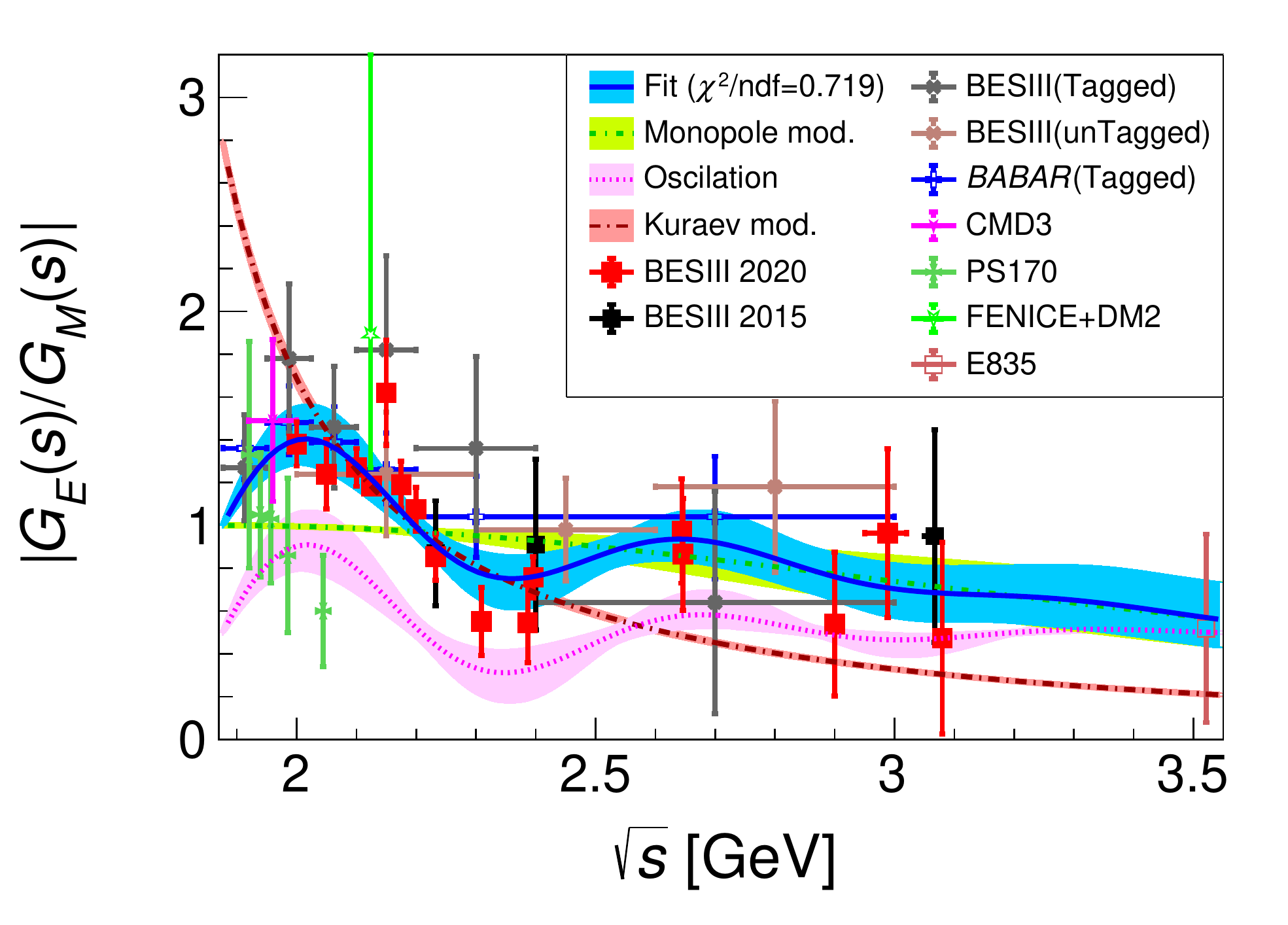}
  \end{overpic}
 }
 \vskip-0.0pt
 \caption{Summary for the ratio $|G_{E}/G_{M}|$ of the proton and a fit to the data (blue solid line and band) according to Eq.~(\ref{Remfit}) with the monopole (green dash-dotted curve and band) and oscillation (magenta dotted curve and band) component (shifted up by 0.5), and the Kuraev model~\cite{Kuraev}  (red dash-dotted curve and band) according to Eq.~(\ref{Remfit2}).
The shown data are the published measurements from BESIII~\cite{xiaorong,BES3untagged,BES3scanLXRC,BES3tagged}, {\textsl{BABAR}}~\cite{introBABAR}, CMD3~\cite{cmd3}, FENICE~\cite{introFENICE,Rinaldo2}, E835~\cite{introE835,Rinaldo2}, PS170~\cite{introPS170}, DM2~\cite{introDM2,Rinaldo2}.
The definition of $\chi^{2}$ is the same as in Figure~\ref{fig_Xsec_low-enegy-eff-QCD}.
}
\label{fig_GEGM_osc}
\end{center}
\end{figure}

Following a similar approach as for $|G_{\text{eff}}|$ in section~\ref{sec_GeffPeriodic}, a fit is performed on the ratio of $|G_{E}|$ and $|G_{M}|$ in the TL region, using a function that consists of a damped oscillation on top of a decreasing monopole part ~\cite{Bianconi3}:
\begin{eqnarray}
\label{Remfit}
|\frac{G_{E}[\omega(s)]}{G_{M}[\omega(s)]}|=\frac{1}{1+\frac{\omega(s)^{2}}{c_{0}}}\{1+c_{1}e^{-c_{2}\omega(s)}\sin[c_{3}\omega(s)]\},
\end{eqnarray}
where $\omega(s)=\sqrt{s}-2m_{p}c^{2}$, and the unitary normalization at the production threshold $|\frac{G_{E}[\omega(4m_{p}^{2})]}{G_{M}[\omega(4m_{p}^{2})]}|=1$ is imposed. 
A fit to the available $|G_{E}/G_{M}|$ results returns the parameters $c_{0}=3.56_{-1.47}^{+4.14}$~GeV$^{2}$, $c_{1}=0.58_{-0.13}^{+0.21}$, $c_{2}=2.17_{-1.05}^{+1.50}$~GeV$^{-1}$ and $c_{3}=9.48_{-0.57}^{+0.46}$~GeV$^{-1}$. 
The results of the fit are shown in Figure~\ref{fig_GEGM_osc} as the blue solid curve and band. The green dash-dotted curve and its band in the same Figure~represent the monopole component, the magenta dotted curve, and its band represent the oscillation component (shifted up by 0.5).
The red dash-dotted curve in Figure~\ref{fig_GEGM_osc} represents the Kuraev model, another form of monopole model~\cite{Kuraev}:
\begin{eqnarray}
\label{Remfit2}
|\frac{G_{E}(s)}{G_{M}(s)}|=\frac{\mu_{p}}{1+\frac{s-4m_{p}^{2}c^{4}}{c_{4}}},
\end{eqnarray}
where $\mu_{p}$ is the proton anomalous magnetic moment ($\mu_{p}=2.79$ in units of Bohr magnetons), $c_{4}=0.74\pm0.03$~GeV from fit.
The unitary normalization at the production threshold $|\frac{G_{E}(2m_{p})}{G_{M}(2m_{p})}|=\mu_{p}$ is imposed. 

\subsection{Individual FFs $|G_{E}|$ and $|G_{M}|$}

Using the definition of the effective FF $|G_{\text{eff}}|$ in Eq.~(\ref{eq_FF2}), the electric and magnetic FF can be expressed in terms of $|G_{\text{eff}}|$ and their ratio $|G_{E}/G_{M}|$~\cite{Bianconi3}:

\begin{eqnarray}
\label{GEfit}
|G_{E}(s)|=|G_{\text{eff}}(s)|\sqrt{\frac{1+\frac{s}{2m_{p}^{2}}}{1+\frac{s}{2m_{p}^{2}|\frac{G_{E}(s)}{G_{M}(s)}|^{2}}}},
 \end{eqnarray}
\begin{eqnarray}
\label{GMfit}
|G_{M}(s)|=|G_{\text{eff}}(s)|\sqrt{\frac{1+\frac{s}{2m_{p}^{2}}}{|\frac{G_{E}(s)}{G_{M}(s)}|^{2}+\frac{s}{2m_{p}^{2}}}}.
\end{eqnarray}

\begin{figure}[htbp]
\begin{center}
\centering
\vskip-0pt
 \mbox{
 \begin{overpic}[width=6.6cm,height=5.0cm,angle=0]{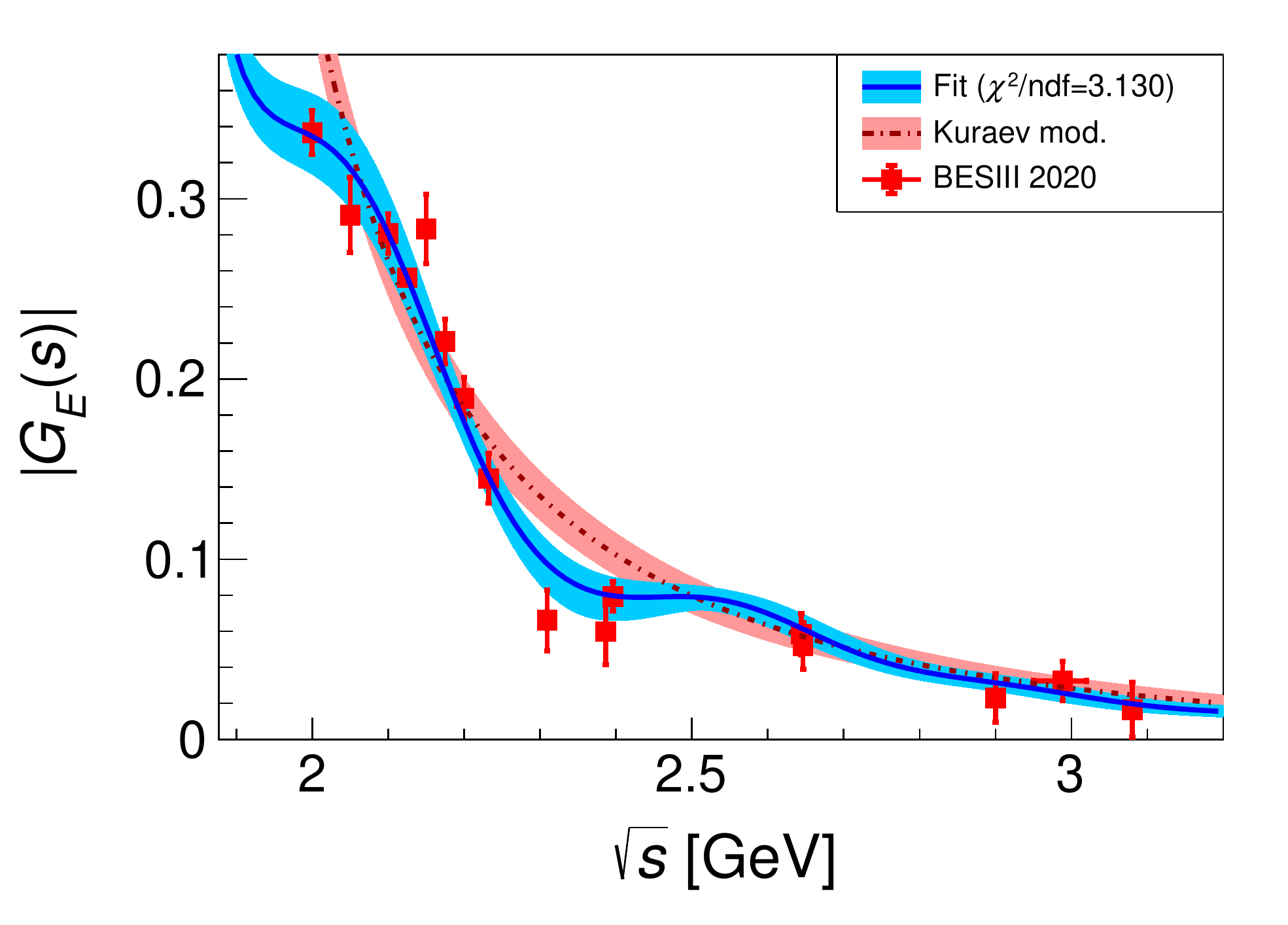}
  \put(0,69){$(a)$}
  \end{overpic}
 \begin{overpic}[width=6.6cm,height=5.0cm,angle=0]{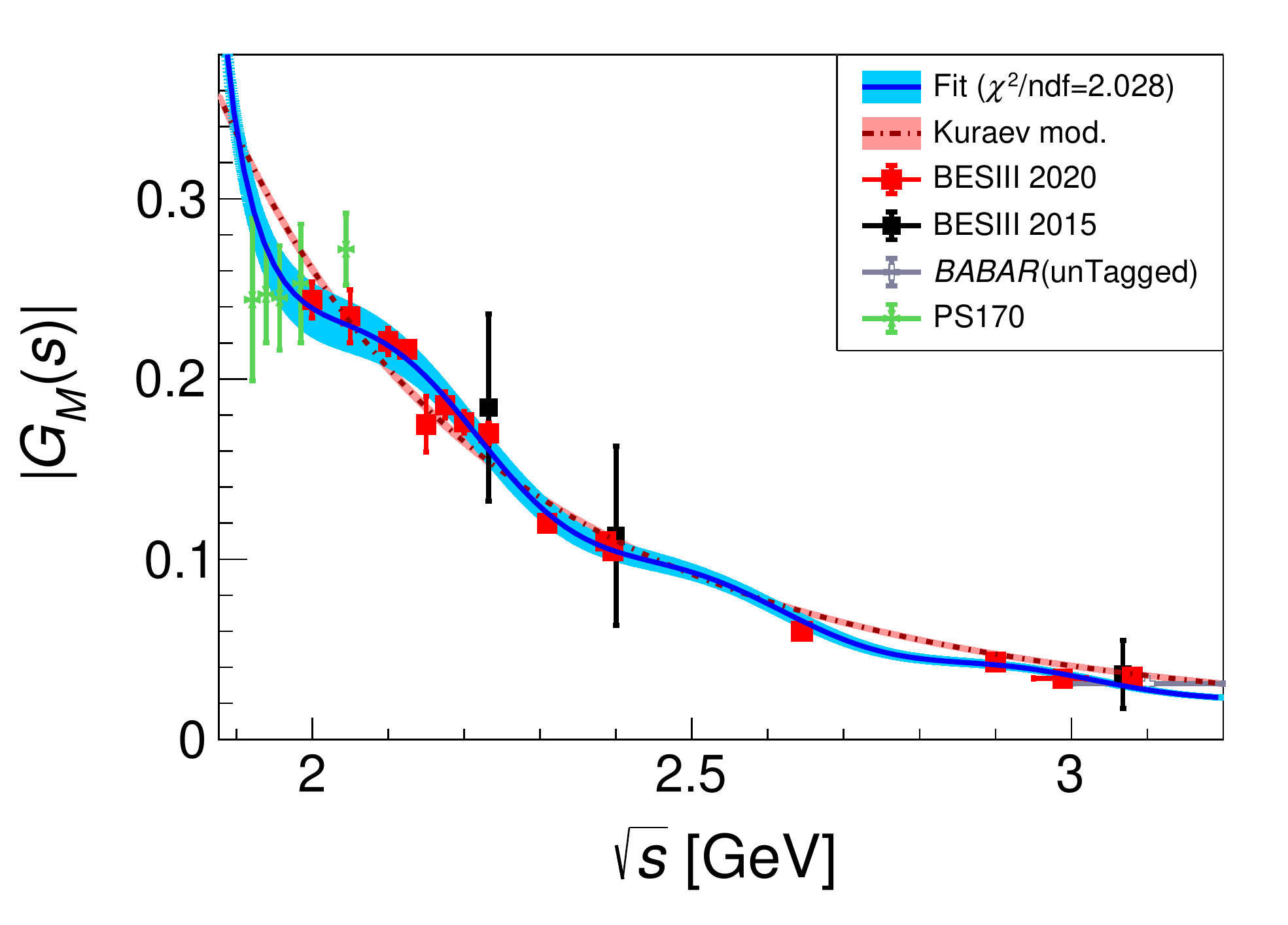}
  \put(0,69){$(b)$}
  \end{overpic} 
  }
 \vskip-0.0pt
 \caption{Summary for (a) the electric FF of the proton $|G_{E}|$ and a fit to the data (blue solid line and band) according to Eq.~(\ref{GEfit}), and the Kuraev model~\cite{Kuraev} Eq.~(\ref{GEKuraev}) (red dash-dotted line and band);
(b) the magnetic FF of the proton $|G_{M}|$ and a fit to the data (blue solid line and band) according to Eq.~(\ref{GMfit}), and the Kuraev model~\cite{Kuraev} Eq.~(\ref{GMKuraev}) (red dash-dotted line and band).
The shown data are the published measurements from BESIII~\cite{xiaorong,BES3untagged,BES3scanLXRC,BES3tagged}, {\textsl{BABAR}}~\cite{introBABAR}, PS170~\cite{introPS170}.
The definition of $\chi^{2}$ is the same as in Figure~\ref{fig_Xsec_low-enegy-eff-QCD}.
}
\label{fig_GE_GM_fit}
\end{center}
\end{figure}

These relations allow to calculate $|G_{E}|$ and $|G_{M}|$ from the fit results obtained for $|G_{E}/G_{M}|$ (Eq.~(\ref{Remfit})) and $|G_{\text{eff}}|$ (Eq.~(\ref{3p-osc})) in the previous section.  
The resulting curves are shown in Figure~\ref{fig_GE_GM_fit}~(a) and~(b). Here, the blue solid curves and bands represent the fit functions according to Eq.~(\ref{GEfit}) and Eq.~(\ref{GMfit}), deduced from the fit results of Figure~\ref{fig_Geff_fit} and Figure~\ref{fig_GEGM_osc}, and the red dash-dotted curves represent the the Kuraev model~\cite{Kuraev}:
\begin{eqnarray}
\label{GEKuraev}
|G_{E}(s)|=\frac{1}{1+\frac{s-4m_{p}^{2}c^{4}}{d_{0}}}\frac{1}{(1+\frac{s-4m_{p}^{2}c^{4}}{d_{1}})^{2}},
\end{eqnarray}
\begin{eqnarray}
\label{GMKuraev}
|G_{M}(s)|=\frac{1}{\mu_{p}(1+\frac{s-4m_{p}^{2}c^{4}}{d_{2}})^{2}},
\end{eqnarray}
where $d_{0}=0.41_{-0.02}^{+0.03}$~GeV$^{2}$, $d_{1}=9.61_{-1.87}^{+2.60}$~GeV$^{2}$ and $d_{2}=2.79\pm0.03$~GeV$^{2}$ are obtained from a fit to the available data.

\subsection{Theoretical estimates of the proton radius}
\begin{figure}[htbp]
\begin{center}
\centering
\vskip-0pt
 \mbox{
 \begin{overpic}[width=10.0cm,height=7.5cm,angle=0]{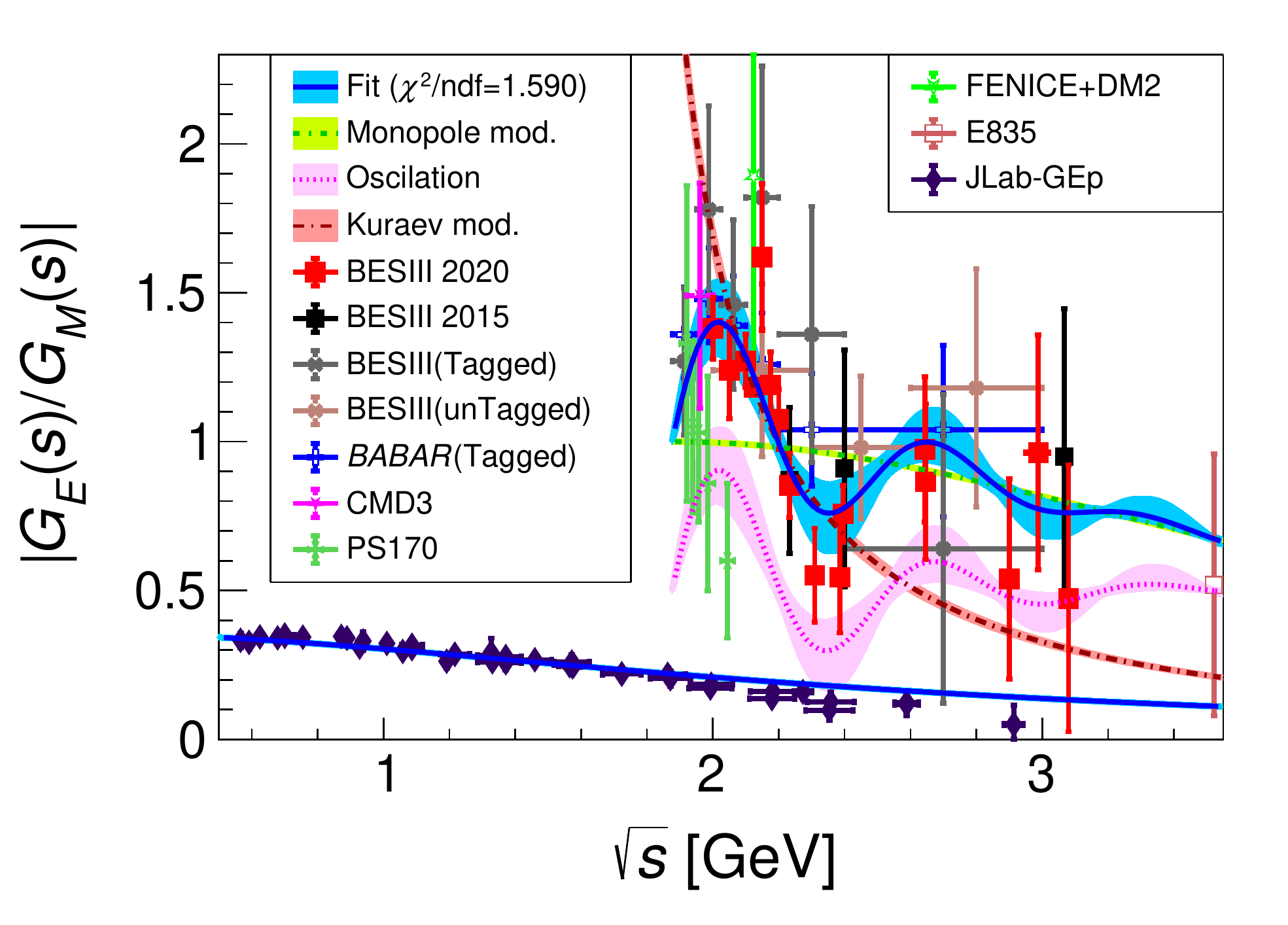}
  \end{overpic}
  }
 \vskip-0.0pt
 \caption{Measured $|G_{E}/G_{M}|$ in the timelike region from BESIII~\cite{xiaorong,BES3untagged,BES3scanLXRC,BES3tagged}, {\textsl{BABAR}}~\cite{introBABAR}, CMD3~\cite{cmd3}, BES~\cite{introBES}, FENICE+DM2~\cite{introFENICE,introDM2,Rinaldo2}, E835~\cite{introE835}, PS170~\cite{introPS170} and in the spacelike region from the GEp Collaboration~\cite{introSL}. The red dash-dotted line and band are monopole-like fits. The definition of $\chi^{2}$ is the same as in Figure~\ref{fig_Xsec_low-enegy-eff-QCD}.}
\label{fig_radius}
\end{center}
\end{figure}

Precise measurements of FFs in the TL region can also be used to improve the theoretical estimates of the proton radius~\cite{Bianconi,Lorenz2}.
A direct comparison of the measured TL and space-like (SL) results of the proton EM FF ratio is shown in Figure~\ref{fig_radius}, together with a simple fit according to the model proposed in Ref.~\cite{Kuraev}:
\begin{eqnarray}
\label{SLKuraev}
|\frac{G_{E}(s)}{G_{M}(s)}|=\frac{1}{\mu_{p}(1+\frac{s}{c_{0}})},
\end{eqnarray}
where $c_{0}$ is the same parameter as in Eq.~(\ref{Remfit}), and is determined to be $c_{0}=5.61_{-1.33}^{+1.37}$~GeV$^{2}$ from the TL-SL joint fit.
The model of Eq.~(\ref{SLKuraev}) allows for a prediction of proton EM FFs by connecting annihilation reactions ($l^{+}l^{-}\xleftrightarrow{}p\bar{p}$) in the TL region to $ep$ elastic scattering in the SL region.
Since there is more data available for SL FFs, they have a larger weight compared to TL FFs. Therefore more TL experimental data over a wider energy range is highly desirable.

\section{Conclusions and Prospect}

Within this review, the progress in the determination and understanding of the proton EM FF in the TL region, both from the experimental and the theoretical point of view, has been highlighted, on the experimental side focusing on results obtained with the energy scan technique. The development of the field, from early pioneering experimental works, for example, at Frascati~\cite{Adone,introFENICE} up to the unprecedented accuracy results from the BESIII experiment~\cite{xiaorong,BES3scanLXRC}, has been outlined, and exciting future prospects such as the PANDA experiment at HESR~\cite{PandaDetector} show that the investigation of proton EM FFs is still a hot topic even 100 years after the discovery of the proton.
\par The first successful investigations of the proton TL EM FFs with the scan technique have been conducted by measurements of the total cross section of $e^{+}e^{-}\xleftrightarrow{} p\bar{p}$ at $e^{+}e^{-}$ or $p\bar{p}$ colliders (illustrated in Figure~\ref{fig_Xsec_result1}), which resulted in a determination of an effective FF $|G_{\text{eff}}|$, assuming $|G_{E}|=|G_{M}|$ (illustrated in Figure~\ref{fig_effFF_result1}). Today, most available measurements only provide this auxiliary quantity, which is extracted under an assumption that does not hold over the whole momentum transfer range. Only recently, experiments have been successful in disentangling the proton EM FFs in the TL region through angular analysis (illustrated in Figure~\ref{fig_cosfit_final}), measuring the ratio of the two FFs $|G_{E}/G_{M}|$ (illustrated in Figure~\ref{fig_Ratio_result}). The high luminosity of the most recent BESIII scan data~\cite{BES3scanLXRC} in addition to a precise knowledge of its integrated luminosity, allowed for a breakthrough in the comprehension of the proton through the individual determination of $|G_{E}|$ and $|G_{M}|$ (illustrated in Figure~\ref{fig_FF_result}).

As shown in the theoretical interpretation of the experimental data for the cross section of the reaction, $e^{+}e^{-}\xleftrightarrow{} p\bar{p}$, pQCD and low-enegy-eff-QCD have been authenticated.
Interesting phenomenology, in particular the superposition of small periodic structures on an otherwise smooth dipole parametrization of $|G_{\text{eff}}|$, has been discovered. Two different approaches currently under discussion to explain these structures have been presented: firstly, the possibility of resonant structures around 2.15 and 2.30~GeV, represented as Breit-Wigner functions on top of the non-resonant pQCD and low-enegy-eff-QCD description (illustrated in Figure~\ref{fig_Xsec_res1}~(a) and (b)), and secondly, a description based on an oscillating function superimposed on the smooth dipole parametrization (illustrated in Figure~\ref{fig_Geff_fit}~(a) and (b)).
Furthermore, a similar periodic behavior for the neutron has been observed (illustrated in Figure~\ref{fig_Geffpn_fit}). Using the oscillation approach, a the conjoint frequency with the proton oscillation of $b^{{\text{osc}}}_{2}=5.05_{-0.09}^{+0.10}$~GeV/$c^{-1}$ has been extracted, with a phase difference to the proton case of $\Delta b^{{\text{osc}}}_{3}=|b^{{\text{osc}}}_{3(p)}-b^{{\text{osc}}}_{3(n)}|=(129_{-10}^{+11})^{\circ}$.
Similar to the oscillations of $|G_{\text{eff}}|$, a periodic behavior of the ratio $|G_{E}/G_{M}|$ has also been observed, which can be extended to $|G_{E}|$ and $|G_{M}|$ (illustrated in Figure~\ref{fig_GEGM_osc} and Figure~\ref{fig_GE_GM_fit}~(a) and (b)). The period of these oscillations can be related to subhadronic scale processes~\cite{Bianconi,Bianconi5}.
We have discussed a faster average decrease in $|G_{E}|$ and $|G_{M}|$, following a similar behavior as in the SL region, which is in agreement with the predictions of Ref~\cite{Kuraev} (illustrated in Figure~\ref{fig_GEGM_osc} and Figure~\ref{fig_GE_GM_fit}~(a) and (b)).
Finally, we have examined the connection of the TL and SL region by a TL-SL joint fit to improve our comprehension of the radius of the proton (illustrated in Figure~(\ref{fig_radius})).

Though a lot of progress has been made to deepen our understanding of the proton structure, most experimental results were reported under the scenario of one-photon approximation. 
In recent years, the scenario of two-photon exchange (TPE) has been re-discussed, starting from a precise measurement of $G_{E}/G_{M}$ ratio by polarized $ep$ scattering experiments~\cite{TPE1,TPE2,TPE3,TPE4}. TPE would manifest in a forward-backward asymmetry in the angular distribution in the TL region.
Precise measurements of such an asymmetry at BESIII are ongoing, bringing valuable new insights into the contribution of TPE to the process.
The ultimate goal in the field of TL nucleon FF’s would be the investigation of the phases between the two complex valued FF’s.
Theoretical efforts in this direction have already been carried out and will be further pursued in the near future. Experimentally, a measurement of the phase would require a measurement of the proton polarization perpendicular to the plane of the incoming particles, which could be achieved by including a polarimeter in the experimental setup of a collider experiment.

In summary, the study of the proton EM FFs in the TL region has come a long way. Measurements with the scan strategy have been the working horse of previous experiments, as well as future prospects such as PANDA. Independent confirmation of recent high precision results, as well as further improvements both on both the precision as well as the measured $q^{2}$ range, will further improve our understanding of the proton's inner structure and dynamics.

\authorcontributions{All authors have read and agreed to the published version of the manuscript.}
\funding{This work is supported in part by National Key Basic Research Program of China under Contract No. 2015CB856700, No. 2020YFA0406403; National Natural Science Foundation of China (NSFC) under Contracts No. 11335008, No. 11375170, No. 11425524, No. 11475164, No. 11475169, No. 11605196, No. 11605198, No. 11625523, No. 11635010, No. 11705192, No. 11735014, No. 11911530140, No. 11950410506, No. 12005219, No. 12035013, No. 12061131003; Joint LargeScale Scientific Facility Funds of the NSFC and CAS under Contracts No. U1532102, No. U1732263, No. U1832103, No. U2032111; China Postdoctoral Science Foundation No. 2021M693097; ERC under contract no. 758462;
European Union Horizon 2020 research and innovation programme under the
Marie Skłodowska-Curie grant agreement no. 645664, 872901, and 894790; German Research Foundation DFG
under contract no. 443159800; Collaborative Research Center CRC 1044,}
\acknowledgments{Authors would like to thank the editors of the special issue of the journal, Dr. Monica Bertani, Prof. Simone Pacetti and Dr. Alessio Mangoni for the organization and helps.
Authors would like to thank Dr. Weiping Wang for their careful reading of this paper.}

\end{paracol}
\reftitle{References}

\end{document}